\documentclass[epj]{svjour}

\usepackage{graphics}
\usepackage{amssymb}
\usepackage{amsmath}
\usepackage{cite}
\usepackage{multirow}
\usepackage{gensymb}
\usepackage{longtable}
\usepackage{booktabs}
\usepackage{array}
\usepackage{anyfontsize} 
\usepackage{ulem}
\usepackage{wrapfig}
\usepackage{xcolor}
\usepackage{cancel}
\usepackage{indentfirst}
\usepackage[figuresright]{rotating}
\usepackage{array}

\definecolor{tit}{rgb}{0.1,0.2,0.4}

\newcommand{\beq}{\begin{eqnarray}}
\newcommand{\eeq}{\end{eqnarray}}
\newcommand{\non}{\nonumber\\ }

\begin{document}
\title{$D_s^{\ast} \to \phi$ helicity form factors and the exclusive weak decays}
\author{Shan Cheng$^{1,2}$, Yao-hui Ju$^{1}$
\and Qin Qin$^{3}$ 
\and Fu-sheng Yu$^{4}$}
%
\mail{Shan Cheng}
\institute{School of Physics and Electronics, Hunan University, 
410082 Changsha, People's Republic of China. 
\and Hunan Provincial Key Laboratory of High-Energy Scale Physics and Applications, 410082, Changsha, China.
\and School of Physics, Huazhong University of Science and Technology, Wuhan 430074, China. 
\and School of Nuclear Science and Technology, Lanzhou University, Lanzhou 730000, China. \non
\non
\email{scheng@hnu.edu.cn, juyaohui@hnu.edu.cn, 2019010193@hust.edu.cn, yufsh@lzu.edu.cn}.
}

\date{Received: date / Revised version: date}

\abstract{
The decay width of the $D_s^{\ast}$ meson is dominated by the electromagnetic mode $D_s^\ast \to D_s \gamma$, 
and it is thus the longest-lived charged vector meson.
In light of this point, we perform the first QCD LCSRs calculation of $D_s^{\ast}\to \phi$ helicity form factors 
and discuss the experiment potential of discovering exclusive $D_s^{\ast}$ weak decays. 
The main result is the partial decay widths, which read as $\Gamma_{D_s^\ast \to l \nu} = 2.44 \times 10^{-12} \, {\rm GeV}$, 
$\Gamma_{D_s^\ast \to \phi l \nu} = (3.28^{+0.82}_{-0.71}) \times 10^{-14} \, {\rm GeV}$, 
$\Gamma_{D_s^{\ast} \to \phi \pi} = ( 3.81^{+1.52}_{-1.33} ) \times 10^{-14} \, {\rm GeV}$ and 
$\Gamma_{D_s^{\ast}\to \phi \rho} = ( 1.16^{+0.42}_{-0.39} ) \times 10^{-13} \, {\rm GeV}$. 
We show that these channels are promising in the near future, serving as the first experimental observation of weak decays of a vector meson, 
and would open up a new playground for precision test of the standard model.}

\PACS{ {13.20.Fc}{Decays of charmed mesons} \and {11.55.Hx}{Sum rules} } 

\maketitle

{\small

\section{Introduction}

The unitarity of Cabibbo-Kobayashi-Maskawa (CKM) matrix is a crucial criterion of the validity of the Standard Model. 
Besides the well known unitarity triangles, which indicate the orthogonality between different rows and columns, 
the CKM unitarity can also be tested by the normalization conditions of individual rows and columns. 
Nowadays the least precisely determinations are
\begin{eqnarray}
&&\vert V_{us} \vert^2 + \vert V_{cs} \vert^2 + \vert V_{ts} \vert^2 = 1.026 \pm 0.022\; , \non
&&\vert V_{cd} \vert^2 + \vert V_{cs} \vert^2 + \vert V_{cb} \vert^2 = 1.025 \pm 0.022 \; , 
\end{eqnarray}
whose uncertainties are both dominated by that of $\vert V_{cs} \vert = 0.987 \pm 0.011$~\cite{PDG2022}. 
The $\vert V_{cs} \vert$ values are typically extracted from semileptonic $D$ decays and leptonic $D_s$ decays, 
and other independent channels, such as weak $D_s^\ast$ decays, are highly anticipated to reduce the uncertainty. 

Weak $D_s^\ast$ decays can also provide a platform for examining the heavy quark symmetry, which is the foundation of 
the heavy quark effective theory~\cite{Neubert:1993mb}. The heavy quark spin symmetry relates the ground-state pseudoscalar 
and vector mesons, {\it e.g.} $D_{(s)}$ and $D_{(s)}^\ast$ mesons. It has been checked through the relation between the semileptonic 
decays ${\bar B} \to D \ell {\bar \nu}$ and ${\bar B} \to D^\ast \ell {\bar \nu}$~\cite{Isgur:1989vq,Nussinov:1986hw,Voloshin:1986dir}, 
where different spin states appear in the final states. The weak $D_s^{\ast +}\to \phi \ell^+ \nu$ decay, together with $D_s^+ \to \phi \ell^+ \nu$, 
will create the first chance to test the heavy quark spin symmetry with  heavy mesons in the initial states. 

Practically, $D_s^{\ast \pm}$ might be the first vector meson whose weak decays will be discovered, 
because it is the longest-lived charged vector meson indicated by the lattice evaluation of the partial width 
of its dominant decay channel ${D_s^\ast \to D_s \gamma}$ \cite{Donald:2013sra}. 
Once the branching ratio of a weak decay channel is measured, it can be used to indirectly determine the total decay width of $D_s^{\ast \pm}$ 
with the theoretical calculation of the weak decay width as an input, 
for which only an experimental upper limit is currently given as $\Gamma_{D_{s}^{\ast}} < 1900 \, \mathrm{keV}$~\cite{PDG2022}.  
Meanwhile, the electromagnetic decay width can also be indirectly determined, 
from which the electromagnetic coupling $g_{D_s^\ast D_s \gamma}$ can be extracted. 
This quantity has been studied by various theoretical approaches (see {\it e.g.} \cite{Li:2020rcg}), but they suffer 
large uncertainties due to the significant destructive interference between radiations of the photon from the charm quark and from the strange quark, 
and also between different QCD power corrections. 
We highlight the recent LCSRs prediction with the complete NLO at twist-1 and twist-2 level \cite{Pullin:2021ebn}, 
the large cancellation between the charm and strange quark contributions are verified 
and the large result $g_{D_s^\ast D_s \gamma} = 0.60^{+0.19}_{-0.18}$ is obtained, which is waiting for the measurement of experiment. 
From another perspective, the $g_{D_s^\ast D_s \gamma}$ coupling is very sensitive to different contributions, 
so the indirect determination from weak $D_s^{\ast \pm}$ decays will subsequently act as an important benchmark to probe the involved dynamics. 

Evaluating the weak $D_s^{\ast}$ decays requires the input of the corresponding heavy-to-light form factors, 
which are basic physical quantities charactering the momentum redistribution of partons after the weak interaction. 
In this paper we study the $D_{s}^\ast \to \phi$ form factors from QCD light-cone sum rules (LCSRs) approach, 
which has been widely applied to calculate form factors in charmed meson 
decays~\cite{Ball:1991bs,Khodjamirian:2000ds,Ball:2006yd,Offen:2013nma,Bediaga:2003hr,Du:2003ja,Wu:2006rd}, 
and this work is its first implementation in a vector-to-vector transition. Different helicity form factors according to 
explicit polarizations of the weak current and the $\phi$ meson are calculated. From the small momentum transfer 
region $0 \leq Q^2 \leq 0.4 \, {\rm GeV}^2$ where the LCSRs predictions are reliable, proper parametrization of the 
form factors is inevitable to extend them to the large region $0.4 \leq Q^2 \lesssim1.2 \, {\rm GeV}^2$.
We employ both the simplified $z$-series expansion formalism~\cite{Bourrely:2008za} and the two-pole parametrization \cite{Becirevic:1999kt}, 
and it turns out that the parametrization scheme does not bring additional considerable uncertainties. 
With the helicity form factors, we obtain the partial decay widths of $D_s^\ast$ weak decays considered here, 
they are $\Gamma_{D_s^\ast \to \phi l \nu} = (3.28^{+0.82}_{-0.71}) \times 10^{-14} \, {\rm GeV}$, 
$\Gamma_{D_s^{\ast} \to \phi \pi} = ( 3.81^{+1.52}_{-1.33} ) \times 10^{-14} \, {\rm GeV}$ and 
$\Gamma_{D_s^{\ast}\to \phi \rho} = ( 1.16^{+0.42}_{-0.39} ) \times 10^{-13} \, {\rm GeV} $. 
These predictions, together with the partial decay width of the leptonic mode $\Gamma_{D_s^\ast \to l \nu} = 2.44 \times 10^{-12} \, {\rm GeV}$, 
promote the experiments to measure the weak decay of a vector meson with the great potential in the near future. 
We remark that the main target of this work is to suggest a feasible measurement of weak decay of vector meson, rather than the precise calculation. 
The accuracy of our prediction of $D_s^\ast \to \phi$ form factors is up to leading order of strong coupling and twist five of two-particle LCDA of $\phi$ meson. 
The contributions form next-to-leading-order (NLO) correction and three-particle LCDAs of $\phi$ meson 
could be accomplished for the study of precise examination after the discovery.

\section{$D_s^\ast \to \phi$ helicity form factors}

We start with the correlation function 
\begin{eqnarray}
F_{\mu a}(q, p_1) 
= i \int d^4x \, e^{iq \cdot x} \, \langle \phi \big\vert T\{ J^{W}_\mu(x), J^{V}_a(0) \} \big\vert 0 \rangle.
\label{eq:correlator}
\end{eqnarray}
In the rest frame of the heavy meson $D_s^{\ast}$, 
the vector current $J^{V}_a = \bar{c} \gamma_a s$ and the weak current $J^{W}_\mu = \bar{s}\gamma_\mu (1-\gamma_5) c$ 
carry momentum $p_{1 a}$ and $q_\mu$, respectively, and hence the momentum of $\phi$ meson is $p_2 = p_1 - q$. 
The kinematics in our convention is arranged by 
\beq
&&p_{1 a} = \left( m_{D_s^\ast}, {\bf 0} \right), \, p_{2 b} = \left( E_2, \bf{p} \right), \, q_\mu = \left( q_0, - {\bf p} \right) \,, \nonumber\\
&&\epsilon_{1 a}(0) = \left(0, 0, 0, 1\right), \,\epsilon_{1 a}(\pm) = \frac{1}{\sqrt 2} \left( 0, \mp 1, - i,0 \right) \,.\nonumber\\
&&\epsilon_{2 b}(0) =  \frac{1}{m_\phi} \left( \vert {\bf p} \vert, 0, 0, E_2 \right), \, 
\epsilon_{2 b}(\pm) = \frac{1}{\sqrt 2} \left( 0, \mp 1, - i, 0 \right) \,, \nonumber\\
&&{\bar \epsilon}_\mu(0) = \frac{1}{\sqrt{q^2}} \left( \vert {\bf p} \vert , 0, 0,  -q_0 \right), \, 
{\bar \epsilon}_\mu(\pm) = \frac{1}{\sqrt 2} \left( 0, \pm 1, -i, 0\right). 
\label{eq:kinematic}
\eeq
We note that the timelike polarisation of leptonic current ${\bar \epsilon}_\mu(t) = (q_0, 0, 0, - \vert {\bf p} \vert)/\sqrt{q^2} \propto q_{\mu}$ 
does not contribute in the semileptonic decaying processes with massless leptons, 
and the other three polarisations, picking up the spin-one part of the off-shell $W$ boson, satisfy $q^\mu {\bar \epsilon}_\mu = 0$. 
Further constraints between these variables can be derived from the kinematical analysis of $1 \to 3$ decaying processes, they are 
\beq
&&2 m_{D_s^\ast} E_2 = m_{D_s^\ast}^2 + m_\phi^2 - q^2, \nonumber\\
&&2 m_{D_s^\ast} q_0 = m_{D_s^\ast}^2 - m_\phi^2 + q^2, \nonumber\\
&&2 m_{D_s^\ast} \vert {\bf p} \vert = \sqrt{\lambda(m_{D_s^\ast}^2, m_{\phi}^2, q^2)}, 
\label{eq:kinematics-constraints}
\eeq
with $\lambda$ being the k\"all${\rm \acute{e}}$n function $\lambda(M_1,M_2,M_3) = M_1^2 + M_2^2 + M_3^2 - 2M_1M_2 - 2M_1M_3 - 2M_2M_3$. 
Multiplying both sides of Eq. (\ref{eq:correlator}) by the polarisation vector of the weak current,  
we can decompose the correlation function in terms of invariant helicity amplitudes, 
\begin{eqnarray}
&~&\bar{\epsilon}^\mu F_{\mu a}(q, p_1) = \sum_{i,j={\bf 0, \pm}} \epsilon^{\ast}_{1 a,  i^\prime} \, F_{ij}(q^2, p_1^2),
\label{eq:Lorentz-decomp}
\end{eqnarray}
here the subscripts $i$, $j$ and $i^\prime = i + j$ denote the polarisation directions of the weak current, $\phi$ meson and vector current, respectively. 

In the view of LCSRs, correlation functions can be formulated in twofold ways, namely, at the quark level and the hadronic level.
Firstly, they can be evaluated directly at the quark-gluon level in the Euclidean momenta space. 
The QCD calculation of Eq. (\ref{eq:correlator}) is carried out with negative $q^2$, 
and the operator product expansion (OPE) is valid for large energies of the final state vector mesons, 
which implies a restriction to not too large momentum transfer squared as $0 \leq \vert q^2 \vert \leq q^2_{{\rm LCSR, max}}$. 
In this region, the operator product of the $c$-quark fields in the correlation function can be expanded near the light cone $x^2 \sim 0$ due to the large virtuality, 
which at leading order reduces to the free quark propagator.

In the QCD evaluation, only the final $\phi$ meson is on shell so that $p_2^2 = (p_1 - q)^2 = m_\phi^2$. 
The OPE calculations obtain the Lorentz decomposition in Eq. (\ref{eq:Lorentz-decomp}) 
where each invariant amplitude can be written in a general convolution of hard functions various LCDAs at different twists~\cite{Ball:2004rg}
\beq
&~&F^{\rm OPE}_{ij}(q^2, (p_2+q)^2) \nonumber\\
&=& \sum_{t} \int_0^1 du \, T^{(t)}_{ij}(u,q^2,(p_2+q)^2) \, \phi^{(t)}(u) \,.
\label{eq:correlator_twistexpansion}
\eeq
The OPE amplitudes is further rewritten in a dispersion integral over the invariant mass of the interpolating heavy meson, 
\begin{eqnarray}
&~&F^{{\rm OPE}}_{ij}(q^2,p_1^2) \nonumber\\
&=& \frac{1}{\pi} \int_{m_c^2}^\infty ds \, \frac{u^2}{[u^2 m_\phi^2 -q^2 + m_c^2]} \, \sum_{n} \frac{\textmd{Im} F^{{\rm OPE}}_{n,ij}(q^2, s)}{u^n [ s-p_1^2 ]^n}, 
\label{eq:OPE-dispersion}
\end{eqnarray}
in which $s \equiv s(q^2,u) = {\bar u} m_\phi^2 + (m_c^2 - {\bar u} q^2)/u$. 
As an example, we present the imaginary parts of the helicity ${\bf 00}$ amplitudes truncated to the third power $n \leq 3$, they are  
\begin{eqnarray}
&~&\frac{1}{\pi} \,{\rm Im} F^{{\rm OPE}}_{1,{\bf 00}}(q^2<0, u) \nonumber\\
&=& \frac{\sqrt{\lambda} m_c f_\phi^\perp m_\phi \phi_2^\perp(u)}{2 m_{D_s^\ast} \sqrt{\vert q^2 \vert}} 
+ \frac{\sqrt{\lambda} (u m_{D_s^\ast}^2 + {\bar u} q^2 ) f_\phi^\parallel \phi_3^\perp(u)}{2 m_{D_s^\ast} \sqrt{\vert q^2 \vert}} \nonumber\\
&-& \frac{\sqrt{\lambda} (m_{D_s^\ast}^2 - q^2 ) f_\phi^\parallel \left[ {\bar \phi}_2^\parallel(u) - {\bar \phi}_3^\perp(u) \right]}{2 m_{D_s^\ast} \sqrt{\vert q^2 \vert}} 
+ \frac{\sqrt{\lambda} f_\phi^\parallel m_\phi^2 {\tilde \psi}_3^\perp(u)}{4 m_{D_s^\ast} \sqrt{\vert q^2 \vert}} \; , 
\label{eq:Im-OPE-L-n1} 
\end{eqnarray}
\begin{eqnarray}
&~&\frac{1}{\pi} \,{\rm Im} F^{{\rm OPE}}_{2,{\bf 00}}(q^2<0, u) \nonumber\\
&=& - \frac{\sqrt{\lambda} \left[ t \lambda + [m_{D_s^\ast}^2 - q^2 + t m_\phi^2](m_{D_s^\ast}^2 - m_\phi^2 + q^2) \right]}{4 m_{D_s^\ast} \sqrt{\vert q^2 \vert}} \nonumber\\ 
&~& \cdot f_\phi^\parallel \, \left({\bar \phi}_2^\parallel(u) - {\bar \phi}_3^\perp(u)\right) \nonumber\\
&-&  \frac{\sqrt{\lambda} \left[ (t+2) m_{D_s^\ast}^2 - m_\phi^2 - (t - 2) q^2 \right]}{4 m_{D_s^\ast} \sqrt{\vert q^2 \vert}} \nonumber\\
&~& \cdot f_\phi^\parallel m_\phi^2 \, \left( \overset{=}{\psi^\parallel_4}(u) + \overset{=}{\phi^\parallel_2}(u) - 2 \overset{=}{\phi^\perp_3}(u) \right) \nonumber\\
&+& \frac{\sqrt{\lambda} \left(m_{D_s^\ast}^2 - m_\phi^2 + q^2 \right)}{4 m_{D_s^\ast} \sqrt{\vert q^2 \vert}} m_c f_\phi^\perp m_\phi \, {\tilde \psi}_3^\parallel(u) \nonumber\\
&+& \frac{\sqrt{\lambda} \left( m_{D_s^\ast}^2 - q^2 \right)}{8 m_{D_s^\ast} \sqrt{\vert q^2 \vert}} 
f_\phi^\parallel m_\phi^2 \, \left( {\bar \phi}_4^\parallel(u) - {\bar \phi}_5^\perp(u) \right)  \nonumber\\
&+& \frac{\sqrt{\lambda}}{2 m_{D_s^\ast} \sqrt{\vert q^2 \vert}} m_c f_\phi^\perp m_\phi^3 \, 
\left( \overset{=}{\psi^\perp_4}(u) + \overset{=}{\phi^\perp_2}(u) - 2 \overset{=}{\phi^\parallel_3}(u) \right) \nonumber\\
&-& \frac{ \sqrt{\lambda} \left( m_{D_s^\ast}^2 + t m_\phi^2 - q^2 \right)}{4 m_{D_s^\ast} \sqrt{\vert q^2 \vert}} 
m_c f_\phi^\perp m_\phi \left( {\bar \psi}_4^\perp(u) - {\bar \phi}_2^\perp(u) \right) \nonumber \\
&-& \frac{ \sqrt{\lambda} \left( u  m_{D_s^\ast}^2 + {\bar u} q^2 \right)}{8 m_{D_s^\ast} \sqrt{\vert q^2 \vert}} f_\phi^\parallel m_\phi^2 \, \phi_5^\perp(u) \nonumber\\
&-& \frac{\sqrt{\lambda}}{16 m_{D_s^\ast} \sqrt{\vert q^2 \vert}} \, f_\phi^\parallel m_\phi^4 \, {\tilde \psi}_5^\perp(u) \,,
\label{eq:Im-OPE-L-n2} \\
&~&\frac{1}{\pi}  \,{\rm Im}  F^{{\rm OPE}}_{3,{\bf 00}}(q^2<0, u) \nonumber\\
&=& - \frac{\lambda^{3/2}}{2 m_{D_s^\ast} \sqrt{\vert q^2 \vert}} m_c f_\phi^\perp m_\phi \,  
\left( \overset{=}{\psi^\perp_4}(u) + \overset{=}{\phi^\perp_2}(u) - 2 \overset{=}{\phi^\parallel_3}(u) \right) \nonumber\\
&+& \frac{f_\phi^\parallel m_\phi^2 \, [ {\bar \phi}^\parallel_4(u)- {\bar \phi}_5^\perp(u) ]}{4} 
\Big[ \frac{m_c^2 \sqrt{\lambda} \left( m_{D_s^\ast}^2 - q^2 \right)} {m_{D_s^\ast} \sqrt{\vert q^2 \vert}} \nonumber\\
&~& + \frac{\sqrt{\lambda}\left[ t \lambda + [m_{D_s^\ast}^2 - q^2 + t m_\phi^2](m_{D_s^\ast}^2 - m_\phi^2 + q^2)\right]}{2 m_{D_s^\ast} \sqrt{\vert q^2 \vert}}  \Big] \nonumber\\
&+& \frac{\sqrt{\lambda} \left[ {\bar u}u\lambda - \left(u m_{D_s^\ast}^2-{\bar u}um_\phi^2+ {\bar u}q^2\right)
\left( m_{D_s^\ast}^2 - m_\phi^2 + q^2\right) \right]}{m_{D_s^\ast} \sqrt{\vert q^2 \vert}} \nonumber\\
&~& \cdot f_\phi^\parallel m_\phi^2 \, \left( \overset{=}{\psi^\parallel_4}(u) + \overset{=}{\phi^\parallel_2}(u) - 2 \overset{=}{\phi^\perp_3}(u) \right) \nonumber\\
&-& \frac{\sqrt{\lambda} }{4 m_{D_s^\ast} \sqrt{\vert q^2 \vert}}  m_c^3 f_\phi^\perp m_\phi^3\, \phi_4^\perp(u) \nonumber\\
&-& \frac{\sqrt{\lambda} \left[ u m_{D_s^\ast}^2 + {\bar u} q^2 \right] }{4 m_{D_s^\ast} \sqrt{\vert q^2 \vert}} m_c^2 f_\phi^\parallel m_\phi^2 \, \phi_5^\perp(u) \nonumber\\
&-& \frac{\sqrt{\lambda}}{8 m_{D_s^\ast} \sqrt{\vert q^2 \vert}} \, f_\phi^\parallel m_\phi^4 m_c^2\, {\tilde \psi}_5^\perp(u)\,,
\label{eq:Im-OPE-L-n3} 
\end{eqnarray}
where $t= 2 u -1$, $\phi_2^{\perp(\parallel)}, \phi_3^{\perp(\parallel)}, {\tilde \psi}_3^{\perp(\parallel)}, \phi_4^\parallel, \psi_4^{\perp(\parallel)}, 
\phi_5^{\perp}, {\tilde \psi}_5^{\perp}$ are the LCDAs of $\phi$ meson at different twists \cite{Ball:2007rt,Ball:2007zt,Bharucha:2015bzk}, 
the auxiliary functions ${\bar \varphi}(u) \equiv \int_0^u du^\prime \varphi(u^\prime)$ and 
$\overset{=}{\varphi}(u) \equiv \int_0^u du^\prime \int_0^{u^\prime} du^{\prime\prime} \varphi(u^{\prime\prime})$ with 
$\varphi \in \{\phi, \psi \}$ satisfy the boundary conditions ${\bar \phi}(0) = {\bar \phi}(1) = 0$ and $\overset{=}{\varphi}(u = 0,1) = 0$, respectively, 
and $\lambda$ refers to the k\"all${\rm \acute{e}}$n function $\lambda(m_{D_s^\ast}^2, m_\phi^2, q^2)$. 
The mass and decay constant are $m_{D_s^\ast} = 2.112 \, {\rm GeV}$, $m_\phi = 1.68 \, {\rm GeV}$ \cite{PDG2022} 
and $f_{D_s^\ast} = 0.274 \, {\rm GeV}$ \cite{Donald:2013sra}. 
The twist four and twist five LCDAs begin to contribute at the subleading power term ($n=2$) 
according to the twist expansion of matrix element from vacuum to $\phi$ meson state. 
The imaginary parts of the other helicity amplitudes (${\bf 0\pm}, {\bf \pm 0}, {\bf \pm\mp}$) are listed in appendix \ref{app:imaginary-amplitudes}. 

When $q^2$ shifts from deeply negative to positive, the typical distance grows between the two currents in Eq.~\eqref{eq:correlator}, 
hence the long-distance quark-gluon interaction begins to form hadrons. 
In this respect, the correlation function can be understood by the sum of contributions from all possible intermediate states with appropriate subtractions. 
The dispersion relation of invariant amplitudes in variable $p_1^2 > 0$ reads 
\begin{align}
F_{ij}(q^2, p^2_1) = \frac{1}{\pi} \int_{m_c^2}^\infty ds \, \frac{{\rm Im} \, F_{ij}(q^2, s)}{s-p_1^2} \,.
\label{eq:hadron-DR}
\end{align}
By inserting a complete set of hadronic states with the quantum number of the ${\bar c}\gamma_a s$ current, 
the spectral function of the ground state is obtained from the optical theorem and written by means of two detached matrix elements   
\begin{align}
\epsilon^\ast_{1a,i'}\rho^{0}_{ij}(q^2) = \bar{\epsilon}^\mu_i \, \langle \phi \big\vert J^{W}_{\mu, j}(x) \big\vert D_s^{\ast} \rangle \langle D_s^\ast \big\vert J^V_a(0) \big\vert 0 \rangle, 
\label{eq:hadron-insertion-2}
\end{align}
in which the latter one is parametrized by the $D_s^*$ decay constant, 
and the former one is written in terms of the $D_s^* \to \phi$ transition form factors associated with orthogonal Lorentz structures \cite{Chang:2019obq,Wang:2007ys}.
\beq
&~&\langle \phi(p_2,\epsilon^\ast_2) \big\vert \bar{s} \gamma_\mu (1-\gamma_5) c \big\vert D_s^{\ast}(\epsilon_1, p_1) \rangle \nonumber\\
&=& (\epsilon_1 \cdot \epsilon_2^\ast) \Big[ p_{1\mu} {\cal V}_1(q^2) - p_{2\mu} {\cal V}_2(q^2) \Big] \nonumber\\
&+& \frac{(\epsilon_1 \cdot q) (\epsilon_2^\ast \cdot q)}{m^2_{D_s^\ast} - m_\phi^2} \Big[ p_{1\mu} {\cal V}_3(q^2) + p_{2\mu} {\cal V}_4(q^2) \Big] \nonumber\\
&-& (\epsilon_1 \cdot q) \epsilon_{2 \mu}^{\ast} {\cal V}_5(q^2) + (\epsilon_2 \cdot q) \epsilon_{1 \mu}^{\ast} {\cal V}_6(q^2) \nonumber\\
&-& i \varepsilon_{\mu\nu\rho\sigma} \epsilon_1^\rho \epsilon_2^{\ast \sigma} \Big[ p_1^\nu {\cal A}_1(q^2) + p_2^\nu {\cal A}_2(q^2) \Big] \nonumber\\
&+& \frac{i \varepsilon_{\mu\nu\rho\sigma} p_1^\rho p_2^\sigma }{m^2_{D_s^\ast} - m_\phi^2} 
\Big[ \epsilon_1^\nu (\epsilon_2^\ast \cdot q) {\cal A}_3(q^2) - \epsilon_2^\nu (\epsilon_1^\ast \cdot q) {\cal A}_4(q^2) \Big] \,,
\label{eq:Ds2phi-ff}
\eeq
here the form factors ${\cal V}_j$ and ${\cal A}_j$ come from the vector and axial-vector currents, respectively. 

We introduce the helicity form factors
\beq
H_{ij} \equiv \bar{\epsilon}^\mu_i \, \langle \phi \big\vert J^{W}_{\mu, j} \big\vert D_s^{\ast} \rangle \,
\label{eq:helicity_ff}
\eeq
and write down the helicity invariant amplitudes as 
\begin{eqnarray}
F_{ij}(q^2 , p_1^2) = \frac{m_{D_s^{\ast}} f_{D_s^\ast}  \, H_{ij}}{m_{D_s^{\ast}}^2 - p_1^2} 
+ \int_{s_0}^\infty ds \, \frac{\rho^{\prime h}_{ij}(q^2, s)}{s - p_1^2} \,. 
\label{eq:hadron-insertion-3} 
\end{eqnarray}
The relations between helicity form factors and Lorentz orthogonal form factors are collected as 
\beq
&~& H_{{\bf 00}}(q^2 > 0) \nonumber\\
&=& \frac{(m_{D_s^\ast}^2 + m_\phi^2 - q^2) \lambda^{1/2}  \left[ - {\cal V}_1(q^2) + {\cal V}_2(q^2) \right]}{4 \sqrt{q^2} m_\phi m_{D_s^\ast}} \nonumber\\
&-& \frac{\lambda^{1/2} \left[ (m_{D_s^\ast}^2 - m_\phi^2 - q^2)  {\cal V}_5(q^2) - (m_{D_s^\ast}^2 - m_\phi^2 + q^2) {\cal V}_6(q^2) \right]}{4 \sqrt{q^2} m_\phi m_{D_s^\ast}} 
\nonumber\\
&+& \frac{\lambda^{3/2}\left[ {\cal V}_3(q^2) + {\cal V}_4(q^2) \right] }{8 \sqrt{q^2} m_\phi m_{D_s^\ast} (m_{D_s^\ast}^2 - m_\phi^2)} ,
\label{eq:helicityff-L} 
\eeq
\beq
&~&H_{{\bf 0 \pm}}(q^2 > 0) \nonumber \\
&=& - \frac{\lambda^{1/2} \left[ {\cal V}_1(q^2) - {\cal V}_2(q^2) \right]}{2 \sqrt{q^2}}  \nonumber\\
&\mp& \frac{\left[(m_{D_s^\ast}^2 - m_\phi^2 + q^2) {\cal A}_1(q^2) + (m_{D_s^\ast}^2 - m_\phi^2 - q^2) {\cal A}_2(q^2) \right] }{2\sqrt{q^2}} \,,
\label{eq:helicityff-LT}\\
&~&H_{{\bf \pm 0}}(q^2 > 0) \nonumber \\
&=& \frac{\lambda^{1/2} {\cal V}_6(q^2) }{2 m_\phi} \pm \frac{\lambda {\cal A}_3(q^2)}{4  m_\phi (m_{D_s^\ast}^2 - m_\phi^2)}   \nonumber \\
&\pm& \left[ \frac{\left( m_{D_s^\ast}^2 + m_\phi^2 - q^2 \right) {\cal A}_1(q^2)}{2 m_\phi} + m_\phi {\cal A}_2(q^2) \right] \,,
\label{eq:helicityff-TL}\\
&~&H_{ {\bf \mp\pm}}(q^2 > 0) \nonumber \\
&=& \frac{\lambda^{1/2}  {\cal V}_5(q^2) }{2 m_{D_s^\ast}} \mp \frac{\lambda  {\cal A}_4(q^2)}{4 m_{D_s^\ast} ( m_{D_s^\ast}^2 - m_\phi^2 )} \nonumber \\
&\mp& \left[ m_{D_s^\ast} \, {\cal A}_1(q^2) + \frac{\left( m_{D_s^\ast}^2 + m_\phi^2 - q^2 \right) {\cal A}_2(q^2) }{2 m_{D_s^\ast}} \right] \,.
\label{eq:helicityff-T}
\eeq

Eqs. (\ref{eq:helicityff-L}-\ref{eq:helicityff-T}) show explicitly the kinematical behavious of the helicity form factors, 
especially at the end-point $q_0^2 = (m_{D_s^\ast} - m_\phi)^2$ 
\beq
H_{{\bf 00}}(q_0^2) = 0\,, \quad
-H_{{\bf 0 \pm}}(q_0^2) = H_{{\bf \pm 0}}(q_0^2) = H_{{\bf \pm \mp}}(q_0^2) \,.
\label{eq:helicity_ff_endpoint}
\eeq
The endpoint relations as shown in Eq. (\ref{eq:helicity_ff_endpoint}) could be understood in terms of rotational symmetry, 
reduction of invariant and the Wigner-Eckart theorem \cite{Hiller:2013cza,Gratrex:2015hna,Hiller:2021zth}, here we take the last one to explain the relations. 
According to the Wigner-Eckart theorem, 
the helicity information in helicity amplitude is only governed by the Clebsch-Gordan (CG) coefficients, 
and the helicity independent dynamics information is absorbed into the matrix elements $M$. 
In our case of $D_s^\ast(\lambda_{cs}) \to \phi(\lambda_\phi) [l \nu_l] (\lambda_q)$ decays, the CG expansion reads as
\beq
H_{\lambda_q {\bar \lambda_\phi} } = C_{\lambda_{cs} \lambda_q {\bar \lambda_\phi}}^{111} M_{111} \,.
\label{eq:helicity-con}
\eeq
The helicity conservation equation $\lambda_{cs} = \lambda_q + {\bar \lambda_\phi}$ with ${\bar \lambda_\phi} = - \lambda_\phi$ is self-evident. 
With taking the CG coefficients $C_{\lambda_{cs} \lambda_\phi \lambda_q}^{j_{cs} j_\phi j_q}$ in the particle data group \cite{PDG2022}, 
we obtain 
\beq
&&H_{00}(q^2_0) \propto C_{000}^{111} = 0 \,, \nonumber\\
&&H_{01}(q_0^2) : H_{0{\bar 1}}(q_0^2)  \propto C_{101}^{111} : C_{{\bar 1}0{\bar 1}}^{111} = -\frac{1}{2} : \frac{1}{2} \,, \nonumber\\
&&H_{10}(q_0^2) : H_{{\bar 1}0}(q_0^2)  \propto C_{110}^{111} : C_{{\bar 1} {\bar 1}0}^{111} = \frac{1}{2} : -\frac{1}{2} \,, \nonumber\\
&&H_{{\bar 1}1}(q_0^2) : H_{1{\bar 1}}(q_0^2)  \propto C_{0 {\bar 1}1}^{111} : C_{01{\bar 1}}^{111} = -\frac{1}{2} : \frac{1}{2} \,, 
\label{eq:endpoint-relation}
\eeq 
which reproduce the end-point relations shown in Eq.(\ref{eq:helicity_ff_endpoint}) if we
consider the replacement $1({\bar 1}) \leftrightarrow +(-)$ between the helicity quantum numbers and the polarization directions. 

Based on the quark-hadron duality, Eqs. (\ref{eq:OPE-dispersion}) and (\ref{eq:hadron-insertion-3}) 
describe the same correlation function from two parallel views,  
so in principle we can solve the helicity form factors by matching the two equations if we know the spectral functions $\rho^{\prime h}_{ij}(s)$. 
We take the semi-local duality to offset the contributions from large $s > s_0$ regions in the two dispersion relation integrals, 
because the magnitude of timelike form factor is close to the spacelike one when the momentum transfer is far away from the resonant state regions,
and they become equal in the QCD limit \cite{Lepage:1980fj,Efremov:1979qk,Chernyak:1983ej}. 
We Borel-transform both sides of the residual contributions below $s_0$ to suppress the pollutions from excited resonant states and continuum spectral, 
and arrive at the sum rules of the helicity form factors, 
\begin{eqnarray}
&~&m_{D_s^{\ast}} f_{D_s^\ast} H_{ij}(q^2) \nonumber \\
&=& \frac{1}{\pi} \int_{m_c^2}^{s_0} ds \, \frac{e^{-(s+m_{D_s^{\ast}}^2)/M^2}}{[u^2(s) m_\phi^2 -q^2 + m_c^2]} \Big[ u(s) \, \textmd{Im} F^{{\rm OPE}}_{1,ij}(q^2, s) \nonumber\\
&~& + \frac{ \textmd{Im} F^{{\rm OPE}}_{2,ij}(q^2, s)}{M^2} + \frac{\textmd{Im} F^{{\rm OPE}}_{3,ij}(q^2, s)}{2 u(s) M^4} \Big] \nonumber\\
&+& \frac{1}{\pi} \frac{e^{-s_0/M^2}}{{[u_0^2 m_\phi^2 -q^2 + m_c^2]}} \Big[ \frac{u_0 \, \textmd{Im} F^{{\rm OPE}}_{2,ij}(q^2, s_0)}{s_0 - q^2} \nonumber\\
&~& + \frac{(1 + x_{s_0}) \textmd{Im} F^{{\rm OPE}}_{3,ij}(q^2, s_0) - u_0 \textmd{Im} F^{\prime {\rm OPE}}_{3,ij}(q^2, s_0) }{2(s_0 - q^2)^2}  
\Big] \,.
\label{eq:correlator-boreltrans}
\end{eqnarray}
Here $u_0$ is the solution of $s_0 = {\bar u} m_\phi^2 + (m_c^2 - {\bar u} q^2)/u$, 
$x_{s_0} \equiv (s_0-q^2)/M^2$ and $\textmd{Im} F^{\prime {\rm OPE}}_{3,ij}(q^2, s_0) = \frac{\partial}{\partial s} \textmd{Im} F^{{\rm OPE}}_{3,ij}(q^2, s) \vert_{s=s_0}$.

The value of Borel mass squared is implied by the internal virtuality of propagator which is smaller than the cutoff threshold value, 
saying $M^2 \sim \mathcal{O}(u m_{D_s^\ast}^2 + {\bar u} Q^2 - u {\bar u} m_\phi^2) < s_0$, 
this value is a litter bit larger than the factorisation scale we chosen at $\mu_f^2 = m^2_{D_s^\ast} - m_c^2 = 1.66^2 \, {\rm GeV}^2$ 
with the quark mass ${\overline m_c}(m_c) = 1.30 \, {\rm GeV}$. 
In practice the selection of Borel mass is actually a compromise between the 
overwhelming chosen of ground state in hadron spectral that demands a small value 
and the convergence of OPE evaluation that prefers a large one, 
which result in a region where $H_{ij}(q^2)$ shows an extremum in $M^2$ \cite{Wang:2007ys,Bharucha:2015bzk}
\beq 
\frac{d}{d(1/M^2)} {\rm ln} H_{ij}(q^2) = 0 \,.
\eeq
The continuum threshold is usually set to close to the outset of the first excited state with the same quantum number as $D_s^\ast$ 
and characterised by $s_0 \approx (m_{D_s^\ast} + \chi)^2$, which is finally determined by considering the maximal stable
evolution of physical quantities on the Borel mass squared. 
From the numerical side, the chose of these two parameters should guaratee the convergence of twist expansion in the truncated OPE calculation 
(high twists contributions are no more than thirty percents) and simultaneously the high energy cutoff in the hadron interpolating 
(the contributions from high excited state and continuum spectral is smaller than thirty percents). 
We finally set them at $M^2 = 4.5 \pm 1.0 \, \rm{GeV}^2$ and $s_0 = 6.8 \pm 1.0 \, {\rm GeV}^2$ in this work. 
The value of Borel mass is a litter bit larger than it chosen in the $D_s \to \pi, K$ transition \cite{Khodjamirian:2000ds}, 
while a litter bit smaller than it chosen in the $D_s \to \phi, f_0(980)$ transition \cite{Bediaga:2003hr}, 
and close to it chosen in the $D_s \to \eta^{\prime}$ transition \cite{Offen:2013nma}. 

\begin{figure*}
\begin{center}
\vspace{2mm}
\resizebox{0.45\textwidth}{!}{
\includegraphics{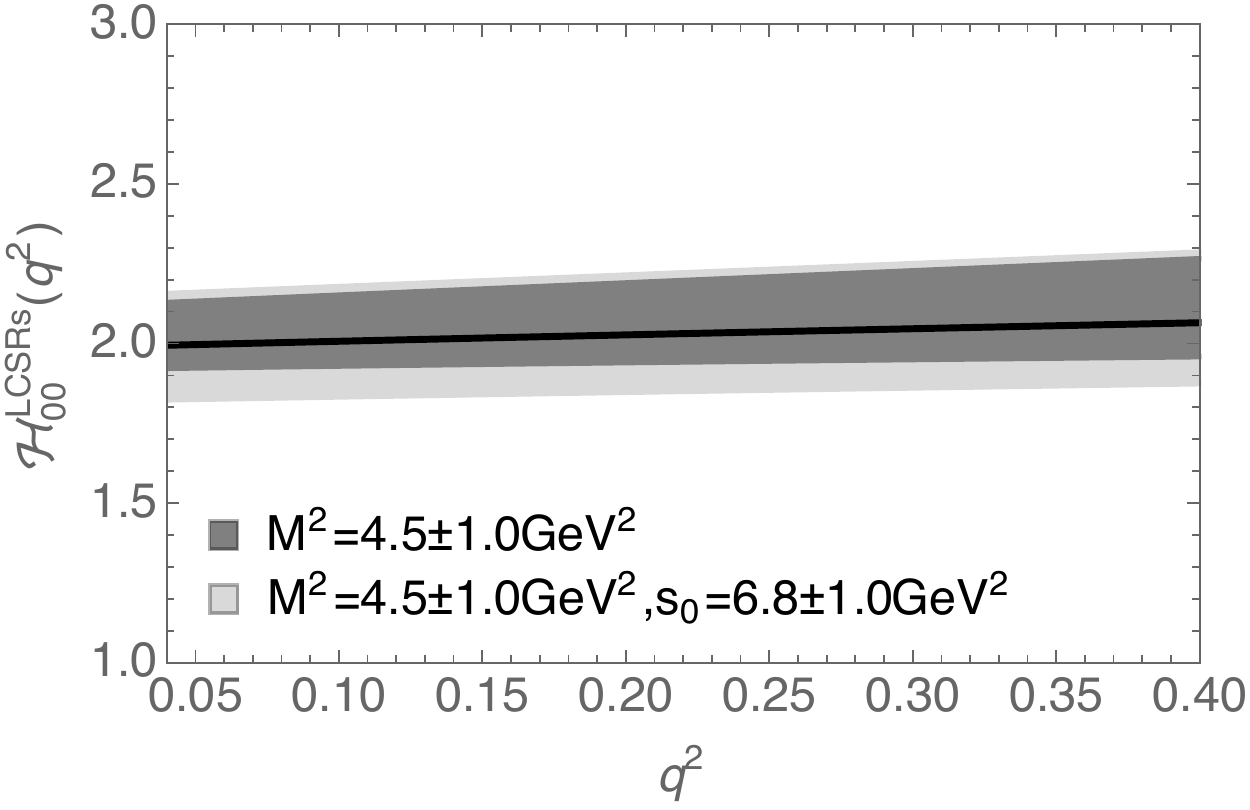}} \non
\vspace{2mm}
\resizebox{0.45\textwidth}{!}{
\includegraphics{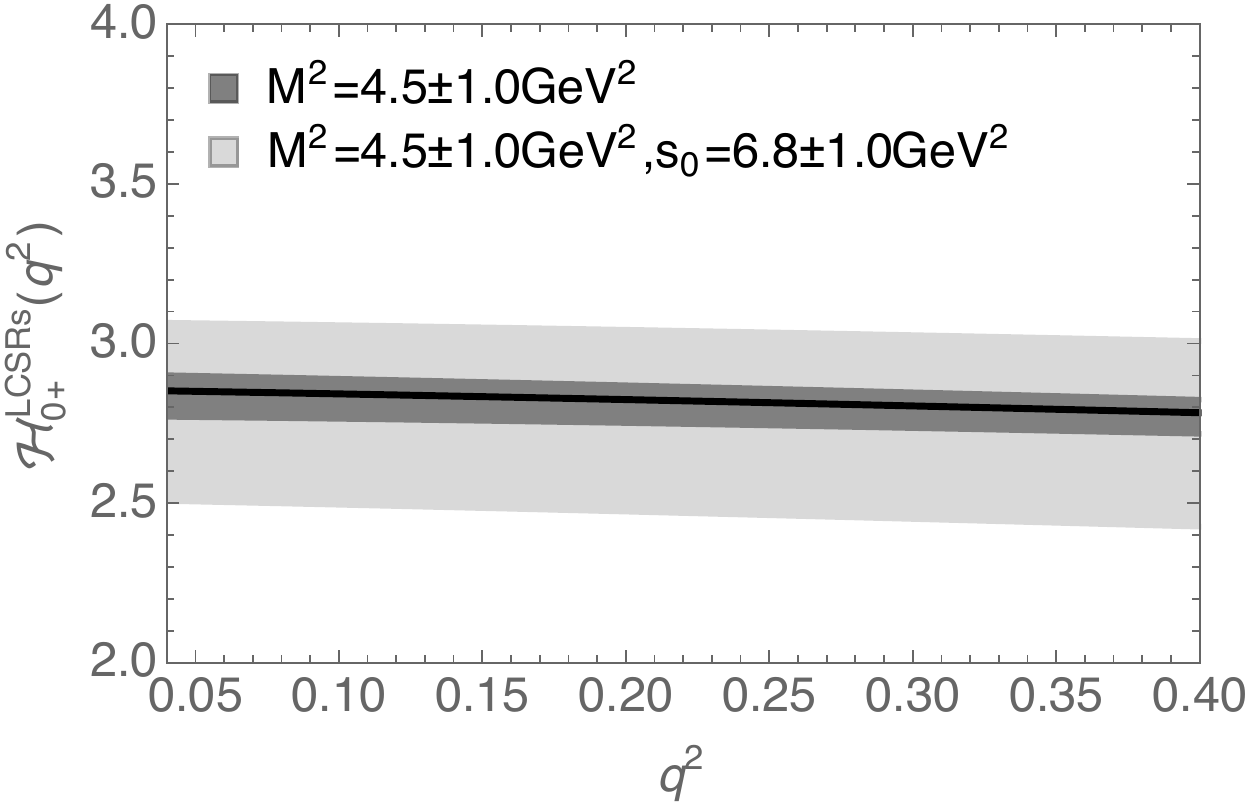}} 
\hspace{8mm}
\resizebox{0.45\textwidth}{!}{
\includegraphics{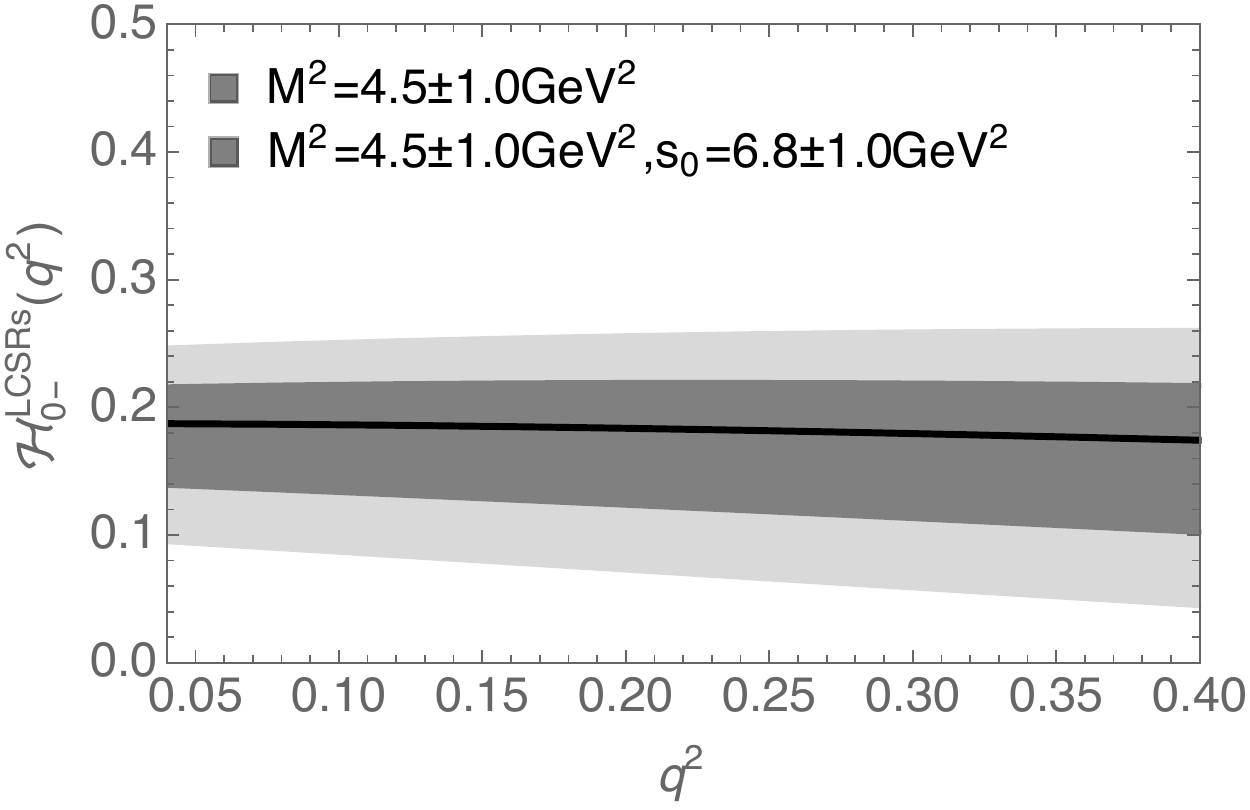}} \non
\vspace{2mm}
\resizebox{0.45\textwidth}{!}{
\includegraphics{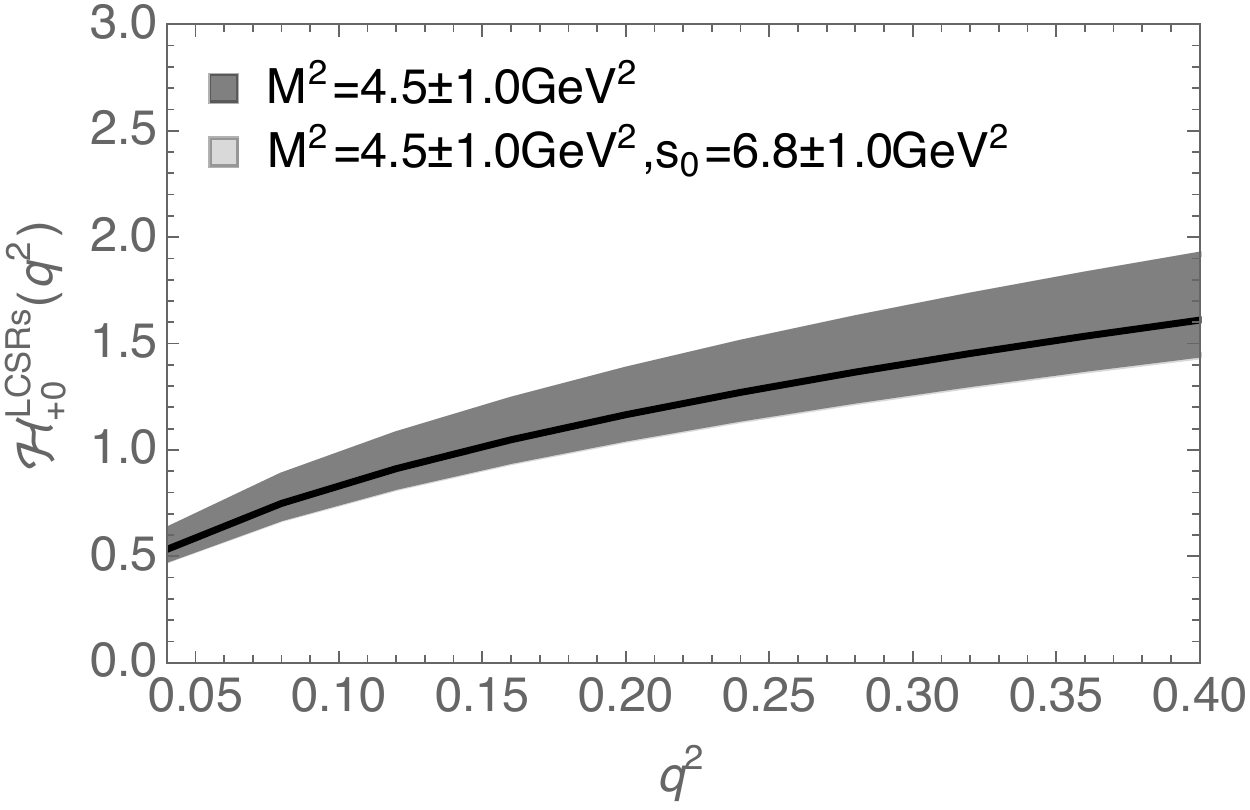}} 
\hspace{8mm}
\resizebox{0.45\textwidth}{!}{
\includegraphics{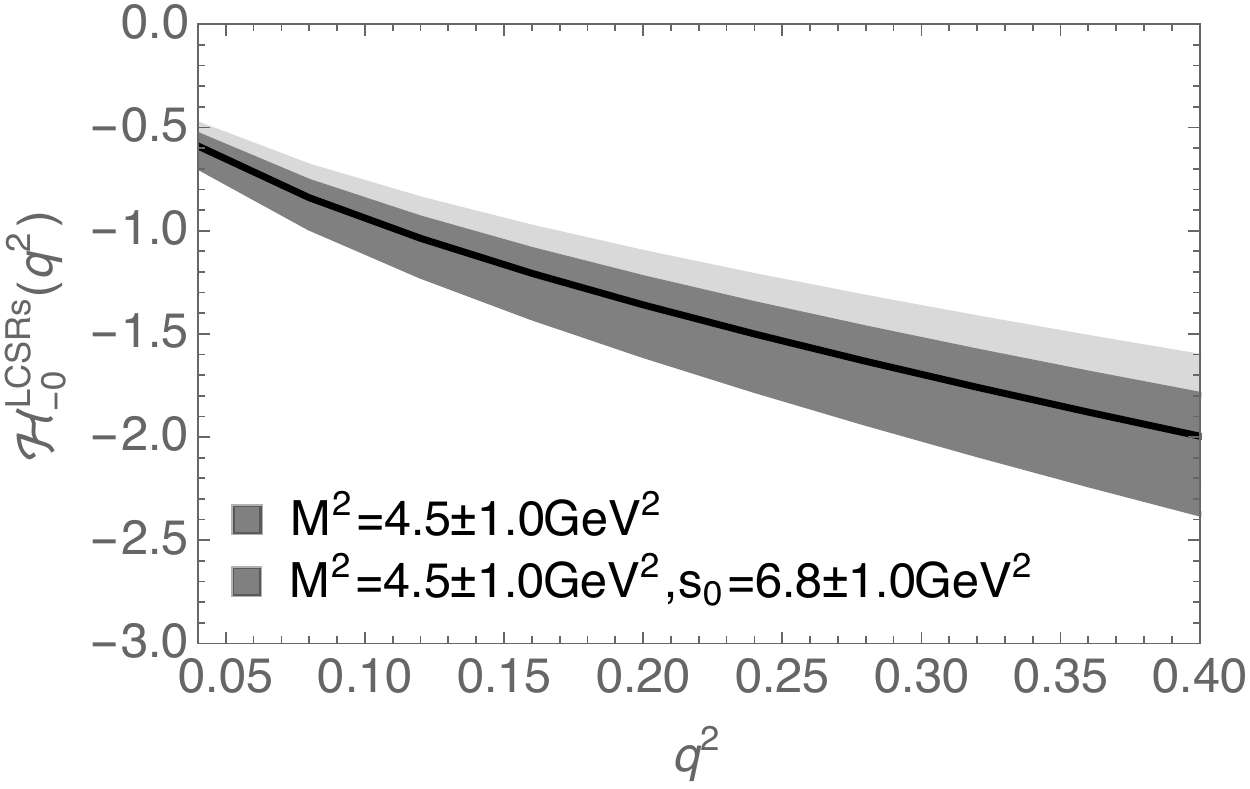}} \non
\vspace{2mm}
\resizebox{0.45\textwidth}{!}{
\includegraphics{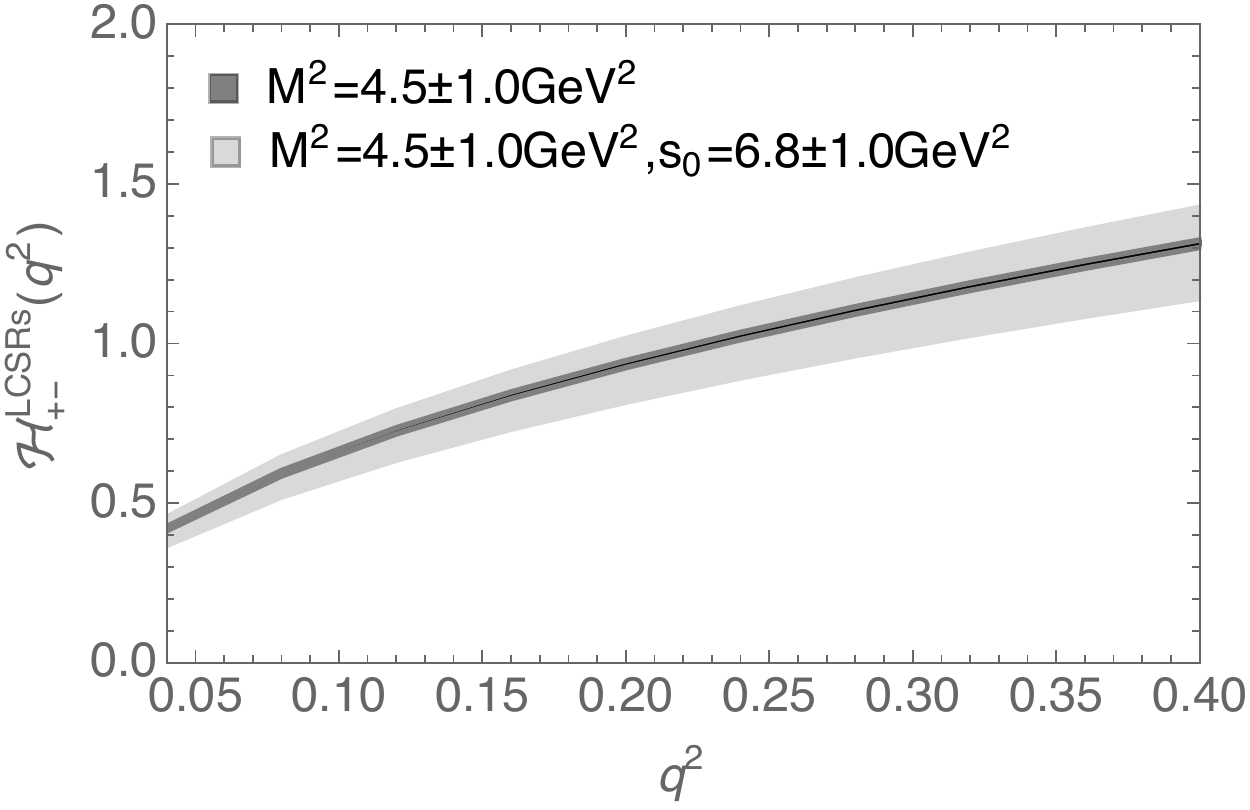}} 
\hspace{8mm}
\resizebox{0.45\textwidth}{!}{
\includegraphics{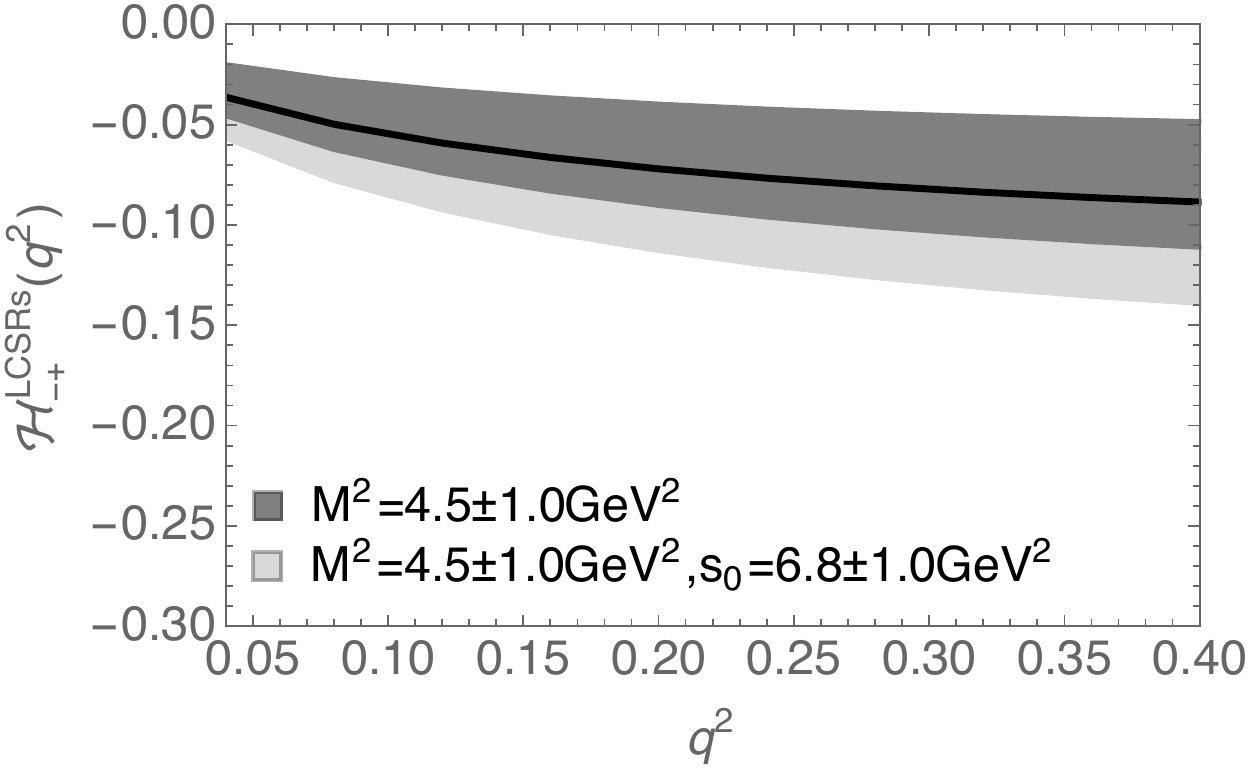}} 
\non
\end{center}
\vspace{-2mm}
\caption{The LCSRs predictions of modified helicity form factors ${\mathcal H}_{ij}(q^2) \equiv \sqrt{q^2} H_{ij}(q^2)$ 
with varying Borel mass in $4.5 \pm 1.0 \, {\rm GeV}^2$ and fixing continuum threshold at $s_0 = 6.8 \, {\rm GeV}^2$ (Gray), 
with varying Borel mass in $4.5 \pm 1.0 \, {\rm GeV}^2$ and continuum threshold in $s_0 = 6.8 \pm 1.0 \, {\rm GeV}^2$.}
\label{fig:1}       
\end{figure*}
\begin{figure*}
\begin{center}
\vspace{2mm}
\resizebox{0.45\textwidth}{!}{
\includegraphics{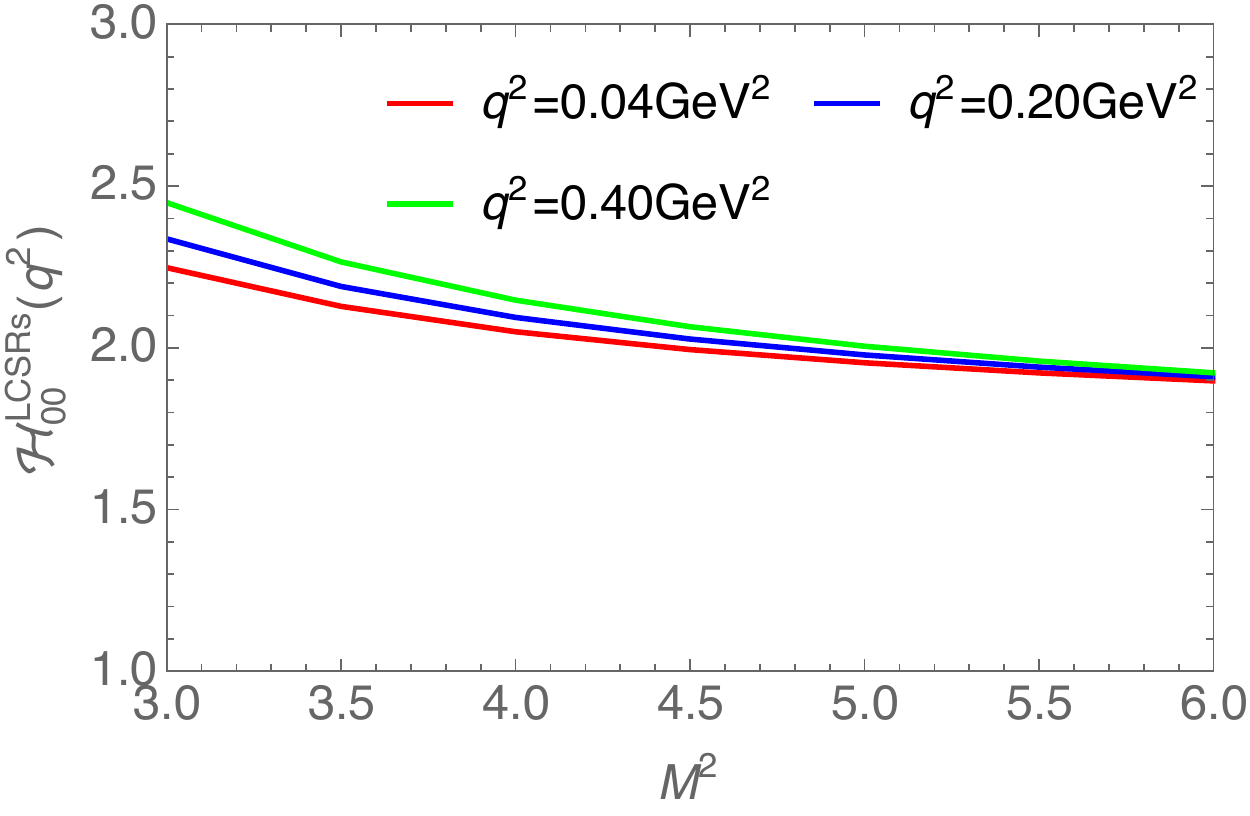}} \non
\vspace{2mm}
\resizebox{0.45\textwidth}{!}{
\includegraphics{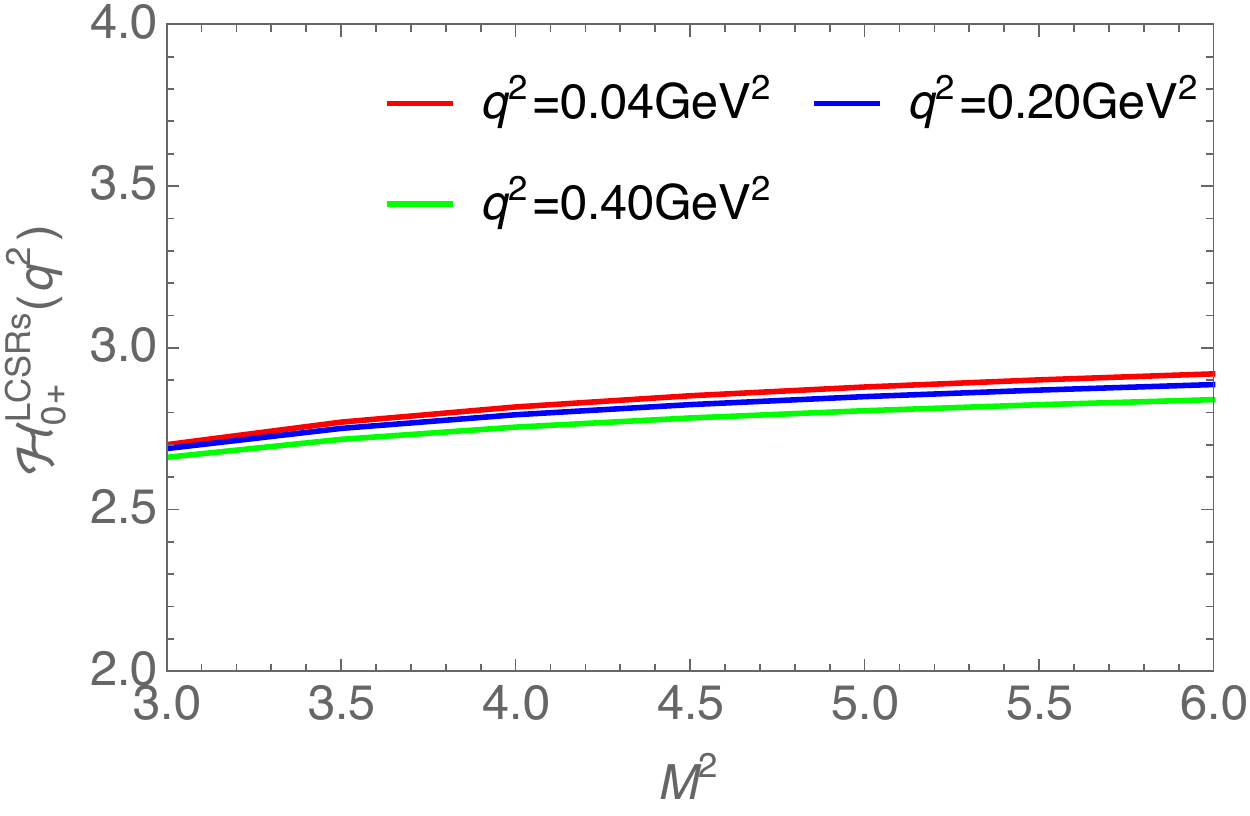}} 
\hspace{8mm}
\resizebox{0.45\textwidth}{!}{
\includegraphics{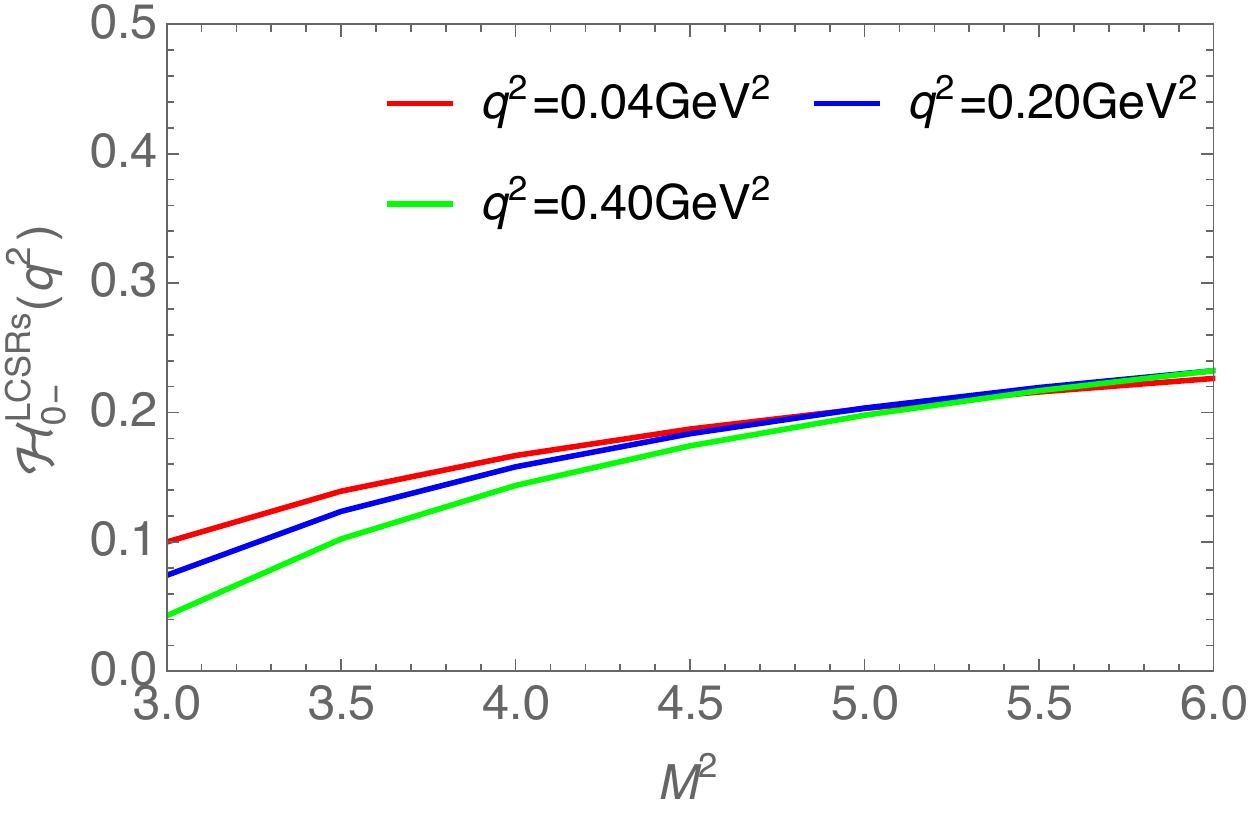}} \non
\vspace{2mm}
\resizebox{0.45\textwidth}{!}{
\includegraphics{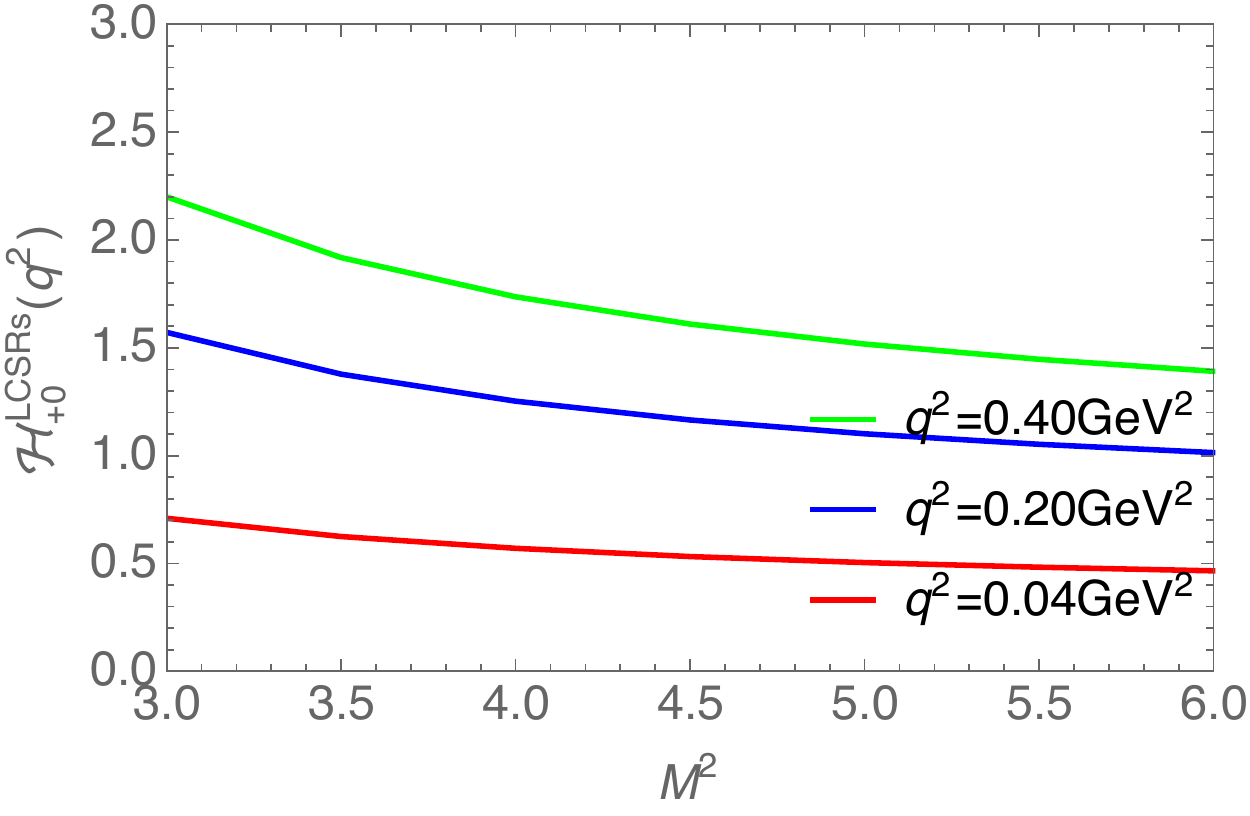}} 
\hspace{8mm}
\resizebox{0.45\textwidth}{!}{
\includegraphics{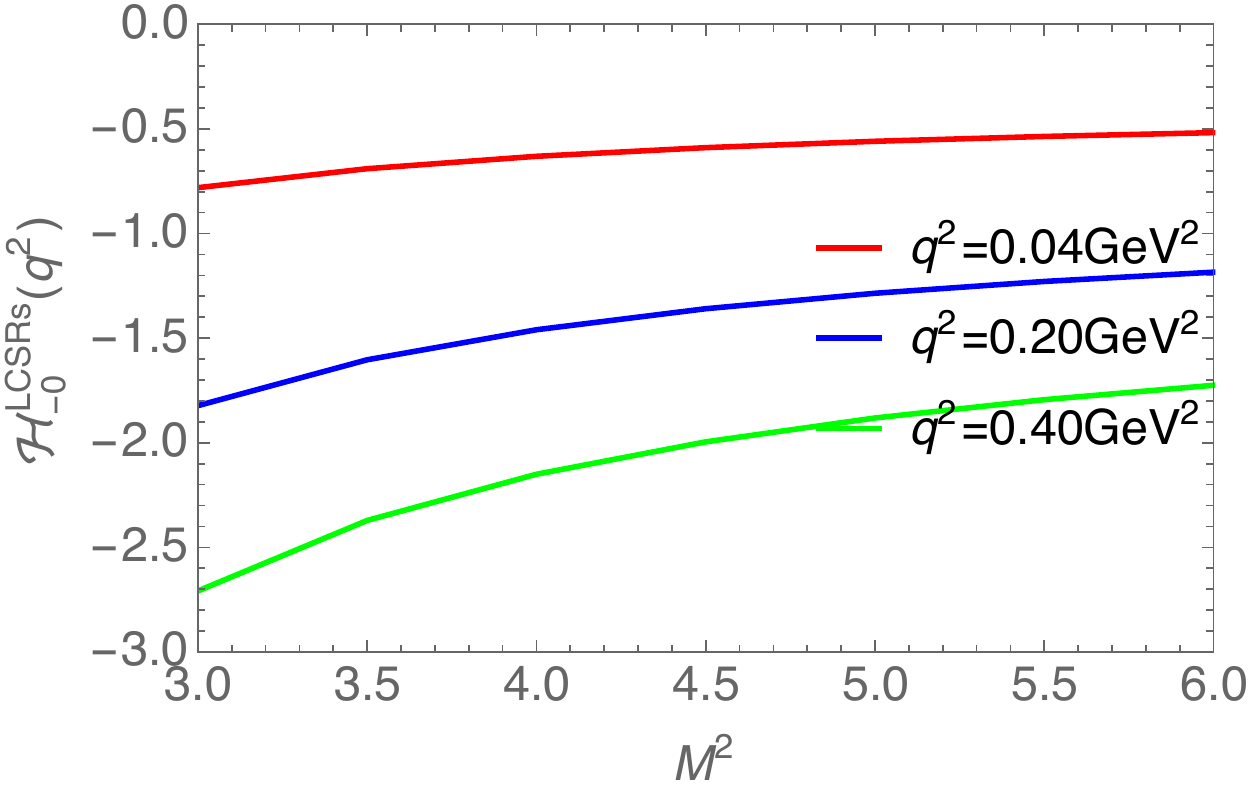}} \non
\vspace{2mm}
\resizebox{0.45\textwidth}{!}{
\includegraphics{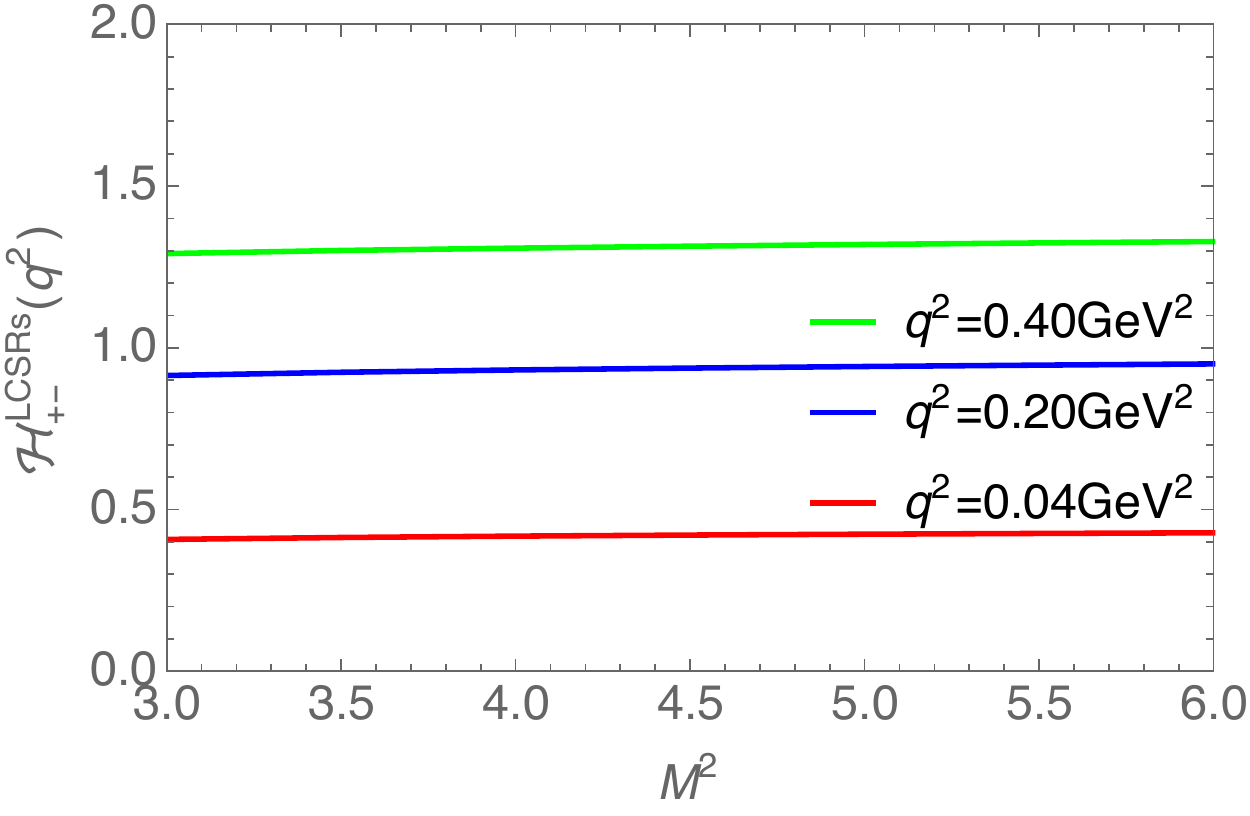}} 
\hspace{8mm}
\resizebox{0.45\textwidth}{!}{
\includegraphics{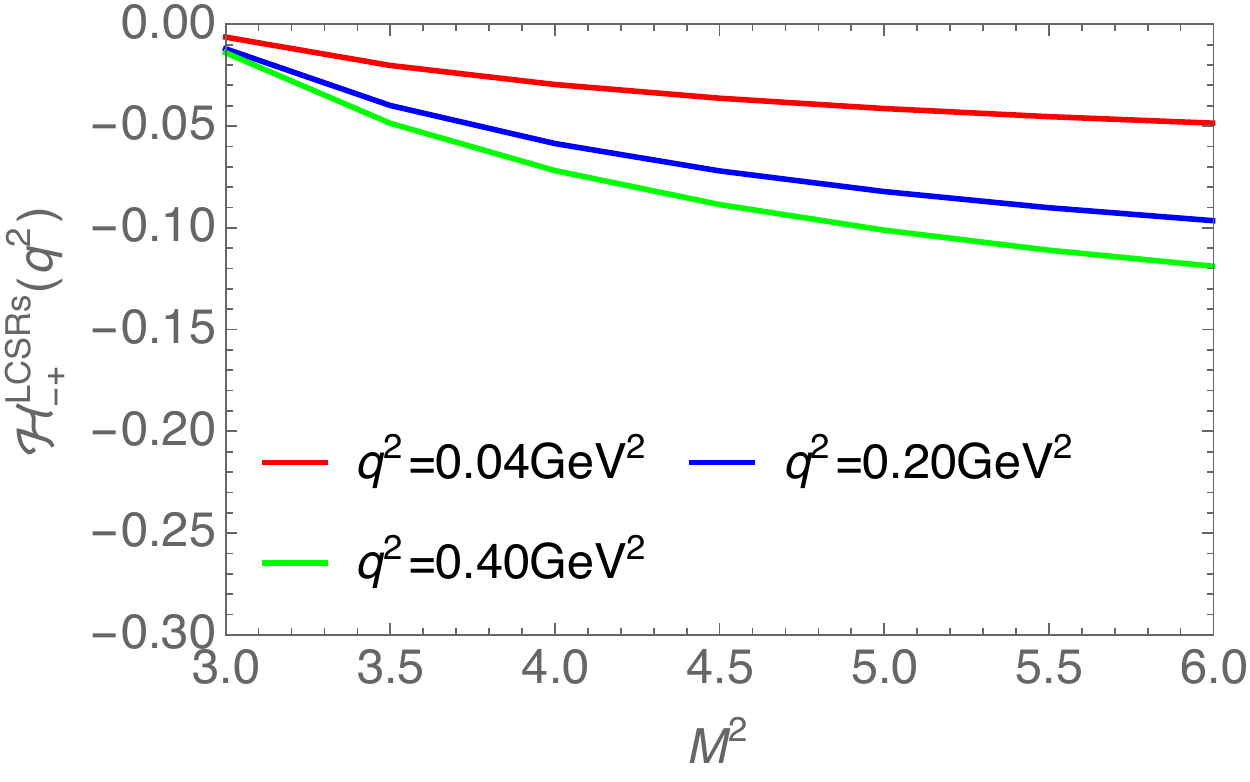}} 
\non
\end{center}
\vspace{-2mm}
\caption{The Borel mass dependence of all seven modified helicity form factors in our considering 
where the continuum threshold is set at $6.8 \, {\rm GeV}^2$. 
Three curves at different momentum transfer points are shown for each form factor.}
\label{fig:2}       
\end{figure*}

\begin{table}
\begin{center}
\caption{The anatomy of the LCSRs uncertainty of helicity form factors ${\mathcal H}_{ij}(q^2)$, 
the center value ({\rm CV}) is obtained by setting $M^2 = 4.5 \, {\rm GeV}^2$ and $s_0 = 6.8 \, {\rm GeV}^2$.}
\vspace{2mm}
\setlength{\tabcolsep}{2mm}{
\begin{tabular}{l | c | c c }
\toprule
${\mathcal H}_{ij}(q^2)$ & ${\rm CV}$&  $M^2\vert^{+1.0}_{-1.0}$ &  $s_0\vert^{+1.0}_{-1.0}$   \nonumber\\
\toprule
${\mathcal H}_{00}(0.04)$ & $1.99$  & $^{-0.07}_{+0.13}$  & $^{+0.09}_{-0.15}$ \nonumber\\
${\mathcal H}_{00}(0.20)$ & $2.03$ & $^{-0.09}_{+0.16}$  & $^{+0.09}_{-0.16}$ \nonumber\\
${\mathcal H}_{00}(0.40)$ & $2.07$ & $^{-0.11}_{+0.20}$  & $^{+0.09}_{-0.16}$ \nonumber\\
\hline
${\mathcal H}_{0+}(0.04)$ & $2.85$ & $^{+0.05}_{-0.08}$ & $^{+0.21}_{-0.33}$ \nonumber\\
${\mathcal H}_{0+}(0.20)$ & $2.82$ & $^{+0.05}_{-0.07}$ & $^{+0.21}_{-0.34}$ \nonumber\\
${\mathcal H}_{0+}(0.40)$ & $2.78$ & $^{+0.04}_{-0.07}$ & $^{+0.22}_{-0.35}$ \nonumber\\
${\mathcal H}_{0-}(0.04)$ & $0.19$  & $^{+0.03}_{-0.05}$ & $^{+0.05}_{-0.08}$ \nonumber\\
${\mathcal H}_{0-}(0.20)$ & $0.18$  & $^{+0.04}_{-0.06}$ & $^{+0.06}_{-0.09}$ \nonumber\\
${\mathcal H}_{0-}(0.40)$ & $0.17$  & $^{+0.04}_{-0.07}$ & $^{+0.07}_{-0.11}$ \nonumber\\
\hline
${\mathcal H}_{+0}(0.04)$ & $0.53$ & $^{-0.05}_{+0.09}$ & $^{+0.01}_{-0.02}$ \nonumber\\
${\mathcal H}_{+0}(0.20)$ & $1.07$ & $^{-0.11}_{+0.21}$ & $^{+0.02}_{-0.04}$ \nonumber\\
${\mathcal H}_{+0}(0.40)$ & $1.61$ & $^{-0.16}_{+0.31}$ & $^{+0.03}_{-0.05}$ \nonumber\\
${\mathcal H}_{-0}(0.04)$ & $-0.59$ & $^{+0.05}_{-0.10}$ & $^{-0.03}_{+0.04}$ \nonumber\\
${\mathcal H}_{-0}(0.20)$ & $-1.36$ & $^{+0.13}_{-0.24}$ & $^{-0.07}_{+0.08}$ \nonumber\\
${\mathcal H}_{-0}(0.40)$ & $-2.00$ & $^{+0.20}_{-0.38}$ & $^{-0.09}_{+0.11}$ \nonumber\\
\hline
${\mathcal H}_{+-}(0.04)$ & $0.42$ & $^{+0.01}_{-0.01}$ & $^{+0.03}_{-0.05}$ \nonumber\\
${\mathcal H}_{+-}(0.20)$ & $0.94$ & $^{+0.01}_{-0.01}$ & $^{+0.08}_{-0.12}$ \nonumber\\
${\mathcal H}_{+-}(0.40)$ & $1.31$ & $^{+0.01}_{-0.01}$ & $^{+0.11}_{-0.17}$ \nonumber\\
${\mathcal H}_{-+}(0.04)$ & $-0.04$ & $^{-0.01}_{+0.02}$ & $^{-0.01}_{+0.01}$ \nonumber\\
${\mathcal H}_{-+}(0.20)$ & $-0.07$ & $^{-0.02}_{+0.03}$ & $^{-0.01}_{+0.02}$ \nonumber\\
${\mathcal H}_{-+}(0.40)$ & $-0.09$ & $^{-0.02}_{+0.04}$ & $^{-0.02}_{+0.03}$ \nonumber\\
\toprule
\end{tabular}}
\end{center}
\label{tab:1}
\end{table}

The tree level LCSRs prediction of modified helicity form factors ${\mathcal H}_{ij}(q^2) \equiv \sqrt{q^2} H_{ij}(q^2)$ 
are depicted in figure \ref{fig:1} where the uncertainties from the Borel mass and the continuum threshold are presented by iteration. 
The Borel mass dependence of these modified helicity form factors are plotted in figure \ref{fig:2}. 
The anatomy of the LCSRs uncertainty are presented in table 1 by taking the result at three momentum transfer points, 
saying $q^2 = 0.04, 0.20$ and $0.40 \, {\rm GeV}^2$. 
It shows that 
\begin{itemize}
\item[(1)] 
Our choice of Borel mass brings $5\%$-$10\%$ uncertainty to the helicity form factor ${\cal H}_{\rm 00}$, 
$10 \%$-$20 \%$ uncertainty to ${\cal H}_{\rm +0}$ and ${\cal H}_{\rm -0}$, 
$20\%$-$30\%$ uncertainty to ${\cal H}_{\rm 0-}$ and ${\cal H}_{\rm -+}$, and less than $5 \%$ uncertainty to ${\cal H}_{\rm 0+}$, 
it almost does not bring uncertainty to the helicity form factor ${\cal H}_{\rm +-}$. 
\item[(2)] 
Our choice of continuum threshold brings another $5\%$-$10\%$ uncertainty to the helicity form factor ${\cal H}_{\rm 00}$, 
$10\%$ uncertainty to ${\cal H}_{\rm 0+}$ and ${\cal H}_{\rm +-}$, $20\%$-$30\%$ uncertainty to ${\cal H}_{\rm -+}$,  
and $30\%$-$40\%$ uncertainty to ${\cal H}_{\rm 0-}$, 
it does not bring additional uncertainty to the helicity form factors ${\cal H}_{\rm +0}$ and ${\cal H}_{\rm -0}$.
\item[(3)]  
The LCSRs uncertainty of form factors ${\cal H}_{\rm 00}$, ${\cal H}_{\rm \pm 0}$ and ${\cal H}_{\rm -+}$ mainly comes from the Borel mass, 
the LCSRs uncertainty of ${\cal H}_{\rm 0-}$ comes equivalently from Borel mass and continuum threshold, 
meanwhile it in form factors ${\cal H}_{\rm 0+}$ and ${\cal H}_{\rm +-}$ mainly arises from the continuum threshold.
\item[(4)]   
These modified helicity form factors have different monotonicities on the two LCSRs parameters, 
for example, ${\cal H}_{\rm 0 \pm}$, ${\cal H}_{\rm -0}$ and ${\cal H}_{\rm +-}$ are monotonically increasing on $M^2$, 
others are monotonically decreasing on $M^2$, 
as shown in figure \ref{fig:2} where the Borel mass dependence of these seven helicity form factors are presented at three different momentum transfer points $q^2 = 0.04, 0.20$ and $0.40 \, {\rm GeV}^2$. 
The magnitudes of all seven modified helicity form factors are all monotonically increasing on the continuum threshold. 
\end{itemize}

\begin{sidewaystable}
\centering
\vspace{8cm}
\caption{The modified helicity form factor ${\mathcal H}_{ij}(q^2) \equiv \sqrt{q^2} H_{ij}(q^2)$ in the large recoiled regions from LCSRs.}
\vspace{2mm}
\setlength{\tabcolsep}{4mm}{
\begin{tabular}{c | c | c c | c c | c c}
\hline
$q^2({\rm GeV}^2)$ & ${\mathcal H}_{00}(q^2)$ & ${\mathcal H}_{0+}(q^2)$ & $ {\mathcal H}_{0-}(q^2)$ 
& ${\mathcal H}_{+0}(q^2)$  & ${\mathcal H}_{-0}(q^2)$ & ${\mathcal H}_{+-}(q^2)$ & ${\mathcal H}_{-+}(q^2)$ \\
\hline
$0.04$ & $1.99^{+0.16+0.32}_{-0.17-0.29}$ & 
$2.85^{+0.21+0.48}_{-0.35-0.47}$ & $0.19^{+0.06+0.09}_{-0.09-0.07}$ & 
$0.53^{+0.09+0.06}_{-0.05-0.06}$ & $-0.59^{-0.11-0.07}_{+0.07+0.07}$ & 
$0.42^{+0.03+0.08}_{-0.05-0.07}$ & $-0.04^{-0.01-0.00}_{+0.02+0.00}$ \\ 
$0.08$ & $2.00^{+0.17+0.31}_{-0.17-0.29}$ & 
$2.85^{+0.21+0.48}_{-0.35-0.47}$ & $0.19^{+0.06+0.09}_{-0.10-0.07}$ & 
$0.75^{+0.13+0.08}_{-0.07-0.08}$ & $-0.84^{-0.15-0.10}_{+0.10+0.10}$ & 
$0.60^{+0.05+0.11}_{-0.08-0.10}$ & $-0.05^{-0.02-0.00}_{+0.03+0.01}$\\
$0.12$ & $2.01^{+0.17+0.31}_{-0.18-0.29}$ & 
$2.84^{+0.22+0.47}_{-0.35-0.46}$ & $0.19^{+0.07+0.08}_{-0.10-0.07}$ & 
$0.91^{+0.16+0.10}_{-0.09-0.10}$ & $-1.04^{-0.19-0.12}_{-0.12-0.12}$ & 
$0.73^{+0.06+0.14}_{-0.09-0.13}$ & $-0.06^{-0.02-0.01}_{+0.03+0.01}$\\
$0.16$ & $2.02^{+0.18+0.30}_{-0.18-0.29}$ & 
$2.83^{+0.22+0.47}_{-0.35-0.46}$ & $0.18^{+0.07+0.08}_{-0.11-0.07}$ & 
$1.05^{+0.19+0.11}_{-0.11-0.11}$ & $-1.21^{-0.22-0.14}_{+0.14+0.13}$ & 
$0.84^{+0.07+0.16}_{-0.11-0.15}$ & $-0.07^{-0.02-0.01}_{+0.04+0.01}$\\
$0.20$ & $2.03^{+0.19+0.30}_{-0.18-0.28}$ & 
$2.82^{+0.22+0.46}_{-0.35-0.46}$ & $0.18^{+0.07+0.08}_{-0.11-0.07}$ & 
$1.07^{+0.21+0.12}_{-0.12-0.12}$ & $-1.36^{-0.25-0.15}_{+0.15+0.15}$ & 
$0.94^{+0.08+0.17}_{-0.12-0.16}$ & $-0.07^{-0.02-0.01}_{+0.04+0.01}$\\
$0.24$ & $2.03^{+0.19+0.29}_{-0.18-0.28}$ & 
$2.82^{+0.22+0.46}_{-0.35-0.45}$ & $0.18^{+0.08+0.08}_{-0.11-0.07}$ & 
$1.27^{+0.23+0.13}_{-0.13-0.13}$ & $-1.50^{-0.28-0.17}_{+0.17+0.16}$ & 
$1.03^{+0.09+0.19}_{-0.13-0.18}$ & $-0.08^{-0.02-0.01}_{+0.04+0.01}$\\
$0.28$ & $2.04^{+0.20+0.29}_{-0.18-0.28}$ & 
$2.81^{+0.22+0.45}_{-0.35-0.45}$ & $0.18^{+0.08+0.08}_{-0.11-0.07}$ & 
$1.37^{+0.25+0.14}_{-0.14-0.14}$ & $-1.63^{-0.31-0.18}_{+0.19+0.17}$ & 
$1.11^{+0.09+0.20}_{-0.14-0.19}$ & $-0.08^{-0.03-0.01}_{+0.05+0.01}$\\
$0.32$ & $2.05^{+0.21+0.29}_{-0.19-0.27}$ & 
$2.80^{+0.22+0.45}_{-0.35-0.45}$ & $0.18^{+0.08+0.08}_{-0.12-0.07}$ & 
$1.45^{+0.27+0.14}_{-0.15-0.15}$ & $-1.76^{-0.34-0.19}_{+0.20+0.19}$ & 
$1.18^{+0.10+0.22}_{-0.15-0.20}$ & $-0.08^{-0.03-0.01}_{+0.05+0.01}$\\
$0.36$ & $2.06^{+0.21+0.28}_{-0.19-0.27}$ & 
$2.79^{+0.22+0.45}_{-0.36-0.44}$ & $0.18^{+0.08+0.08}_{-0.12-0.07}$ & 
$1.53^{+0.29+0.15}_{-0.16-0.15}$ & $-1.88^{-0.36-0.21}_{+0.21+0.20}$ & 
$1.25^{+0.11+0.23}_{-0.16-0.21}$ & $-0.09^{-0.03-0.01}_{+0.05+0.01}$\\
$0.40$ & $2.07^{+0.22+0.28}_{-0.19-0.27}$ & 
$2.78^{+0.22+0.44}_{-0.36-0.44}$ & $0.17^{+0.09+0.08}_{-0.12-0.07}$ & 
$1.61^{+0.31+0.15}_{-0.17-0.16}$ & $-2.00^{-0.39-0.22}_{+0.23+0.21}$ & 
$1.31^{+0.11+0.24}_{-0.17-0.22}$ & $-0.09^{-0.03-0.01}_{+0.05+0.01}$\\
\hline
\end{tabular}}
\label{tab:2}
\end{sidewaystable}

In table \ref{tab:2}, we show the LCSRs prediction of modified helicity form factors at the fixed momentum transfer points, 
saying from $0.04$ to $0.4 \,{\rm GeV}$ with the step $0.04 \, {\rm GeV}$. 
The first uncertainties come from the LCSRs parameters $M^2$ and $s_0$ which is added by the quadratic sum. 
In order to estimate the effect from the missing radiative corrections, 
we vary the charm quark mass in the intervel  ${\overline m_c}(m_c) = 1.30 \pm 0.10 \, {\rm GeV}$ 
and regard this possible NLO effect as the second uncertainty. 
The facotrization scale is then varied in $\mu_f = 1.66 \pm 0.08 \,{\rm GeV}$ correspondingly.
It brings about $10\%$ additional uncertainty to the modified helicity form factors ${\cal H}_{\pm 0}$ and ${\cal H}_{-+}$, 
$15\%$ additional uncertainty to ${\cal H}_{00}$ and ${\cal H}_{0+}$, 
$20\%$ additional uncertainty to ${\cal H}_{+-}$, 
meanwhile $40 \%$ additional uncertainty to ${\cal H}_{0-}$. 
We examined the affect to the Borel mass determination from the quark mass variation 
and found that $M^2 = 4.5 \pm 1.0 \, {\rm GeV}^2$ is still the optimal choice. 
We do not present the uncertainty from the nonperturbative parameters in $\phi$ meson LCDAs as shown in table \ref{tab:4}, 
since the decay constants from the lattice evaluation almost do not bring additional uncertainty and the 
the uncertainty associated to strange quark mass is less than two percent. 

We remark again that the main target of this work is to discuss a feasible measurement of the weak decay of vector meson, 
so the staring point for the calculation is the multiplied correlation function in Eq. (\ref{eq:Lorentz-decomp}) which deduces to the helicity form factors. 
Moreover, what we have indeed calculated is the seven helicity form factors involved in the semileptonic weak decay, 
and hence we can not obtain the ten orthogonal Lorentz form factors corresponding to the correlation function in Eq. (\ref{eq:correlator}) 
by a linearly variation. 
But their relations as shown in Eqs. (\ref{eq:helicityff-L}-\ref{eq:helicityff-T}) provide some constraints to deduce the orthogonal Lorentz form factors. 
For example, the (modified) helicity form factors at the full recoiled point $q^2=0$ are 
\beq
&& {\mathcal H}_{00}(0) = 1.99^{+0.15+0.32}_{-0.17-0.30}\,,\nonumber\\
&& {\mathcal H}_{0+}(0) = 2.86^{+0.21+0.48}_{-0.35-0.48}\,, \quad {\mathcal H}_{0-}(0) = 0.19^{+0.05+0.09}_{-0.09-0.07}\,,\nonumber\\
&&H_{+0}(0) = 2.67^{+0.47+0.31}_{-0.26-0.29}\,, \quad H_{-0}(0) = -2.92^{-0.53-0.35}_{+0.33+0.32}\,,\nonumber\\
&&H_{+-}(0) = 2.11^{+0.17+0.40}_{-0.27-0.28}\,, \quad H_{-+}(0) = -0.19^{-0.06-0.01}_{+0.11+0.03}\,,  
\eeq
from which we can deduce the center values of several orthogonal Lorentz form factors as 
${\cal V}_1(0) - {\cal V}_2(0) = -1.86$, ${\cal V}_5(0) = 2.46$, ${\cal V}_6(0) = -0.26$, and ${\cal A}_1(0) + {\cal A}_2(0) = -1.63$.
These value can be compared with the result obtained from other approaches such as the light-front quark model \cite{Chang:2019obq}, 
and in fact they show a good consistence after considering the different definitions of ${\cal V}_{i=1-6}$ and ${\cal A}_{i=1-4}$ 
in Eq. (\ref{eq:Ds2phi-ff}) here and Eqs. (2.1,2.2) there. 

To extrapolate to the whole kinematic region $[0, q_0^2]$, 
we adopt the SSE parameterisation \cite{Bourrely:2008za} which is required not only 
to reproduce the result obtained from LCSRs calculation in the lower interval $[0, q^2_{{\rm LCSR, max}}]$ with good accuracy, 
but also to provide an extrapolation to the up interval $[q^2_{{\rm LCSR, max}}, q_0^2]$ with the expected analytical properties of the helicity form factors. 
For the maximal momentum transfer squared where LCSRs is still applicable, 
we take it at $m_c^2 - 2 m_c \chi \sim 0.4 \, {\rm GeV}^2$ with $\chi \sim 0.5 \, {\rm GeV}$ being a typical hadronic scale,
as what have been done in $D_{(s)}$ decays \cite{Khodjamirian:2000ds,Ball:2006yd}. 
We truncate the simplified $z$-series expansion after the linear term for the Lorentz orthogonal form factors ${\cal F}_{i} = {\cal V}_{1-6}, {\cal A}_{1-4}$, 
\beq
&&{\cal F}_i(q^2>0) = \frac{a_{{\cal F}_i}}{1-q^2/m_{D1}^2} \Big\{1 + b_{{\cal F}_i} \, [ z(q^2) - z(0) ]  \Big\} \,, 
\label{eq:z-para}
\eeq 
the quadratic term is checked could be negligible here. 
In the expansion, $1/(1-q^2/m_{D1}^2)$ denotes the simple pole corresponding to the lowest-lying resonance in the $D_s^\ast \phi$ spectrum with 
$m_{D1} = 2.77 \, {\rm GeV}$ \cite{PDG2022}, 
$a_{{\cal F}_i} \equiv {\cal F}_i(0)$ indicates the normalization conditions. 
The SSE formula bases on a rapidly converging series 
\beq
&&z(q^2) = \frac{\sqrt{t_+ - q^2} - \sqrt{t_+ - t_0}}{\sqrt{t_+ - q^2} + \sqrt{t_+ - t_0}} \,
\label{eq:z-para-z} 
\eeq
with $t_\pm \equiv (m_{D_s^\ast} \pm m_\phi)^2$ and $t_0 \equiv t_+ (1- \sqrt{1-t_-/t_+})$. 

We parameterize the helicity form factors by considering their general kinematical behavious in Eqs. (\ref{eq:helicityff-L}-\ref{eq:helicityff-T}) 
and their end-point relations in Eq. (\ref{eq:helicity_ff_endpoint}), their expressions are 
\beq
&~&{\mathcal H}^{\rm SSE}_{ij}(q^2) \equiv \sqrt{q^2} H_{ij}(q^2) \nonumber \\
&=& {\cal K}_{ij}(q^2) \frac{1 + b_{ij} z^\prime(q^2)  }{\left( 1 - q^2/m^2_{D1} \right)} 
\Big[ \frac{a_{ij}^1 \lambda^{3/2} }{m_{D_s^\ast} m_\phi (m_{D_s^\ast}^2 - m_\phi^2 )} \nonumber\\
&~& \hspace{2cm} + \frac{ a_{ij}^2  \lambda}{(m_{D_s^\ast}^2 - m_\phi^2 )} + a_{ij}^3 \lambda^{1/2} \Big] \nonumber\\
&+& \kappa_{ij} q_0 \left[ m_{D_s^\ast} \frac{1 + b_1^{\bf ED} z^\prime(q^2)  }{\left( 1 - q^2/m^2_{D1} \right)} 
+ m_\phi \frac{1 + b_2^{\bf ED} z^\prime(q^2)  }{\left( 1 - q^2/m^2_{D1} \right)} \right] \,. 
\label{eq:aDssphi-q2}
\eeq
Here $z^\prime(q^2) \equiv z(q^2) - z(0)$ and $q_1^2 \equiv q_0^2 - q^2$. 
The kinematical functions/factors read as 
\beq
&&{\cal K}_{\bf 00}(q^2) = \frac{\lambda^{1/2} }{m_{D_s^\ast} m_\phi}, \quad \kappa_{\bf 00} = 0 \,;  \nonumber\\
&&{\cal K}_{ij \neq \bf{ 00}} = 1 \,;  \quad \kappa_{\bf 0 \mp}(q^2) = \kappa_{\bf \pm 0}(q^2) = \kappa_{\bf \pm \mp}(q^2) = \pm 1 \,; \nonumber\\
&&a_{0 \pm}^{1} = a_{\pm 0 }^{1} = a_{\pm\mp }^{1} = 0 \,, \quad a_{0 \pm}^{2} = 0 \,.
\eeq
We can see that the terms in the third line on the right hand side give the result of form factors ${\cal H}_{ij \neq \bf{00} }$ 
at the kinematical end-point with the universal parameters $ b_{1,2}^{\bf ED}$, 
and the general terms in the first two lines hint the end-point constraint of ${\cal H}_{\bf{00} }(q_0^2) = 0$.

\begin{table*}
\begin{center}
\caption{The SSE parameters of the helicity form factors ${\mathcal H}_{ij}(q^2)$.}
\vspace{2mm}
\setlength{\tabcolsep}{2mm}{
\begin{tabular}{l | r | r r | r r | r r }
\toprule
Para. & ${\mathcal H}_{00}(q^2)$ &  ${\mathcal H}_{0+}(q^2)$ &  ${\mathcal H}_{0-}(q^2)$ & 
${\mathcal H}_{+0}(q^2)$ & $ {\mathcal H}_{-0}(q^2)$ & ${\mathcal H}_{+-}(q^2)$ & ${\mathcal H}_{-+}(q^2)$  \nonumber\\
\toprule
$a_{ij}^1$ & $0.70^{+0.21}_{-0.05}$  &  $0$ & $0$  & $0$ & $0$ &  $0$ & $0$  \nonumber\\
$a_{ij}^2$ & $-2.55^{-0.79}_{+0.58}$ & $0$ & $0$  & $-2.00^{-0.76}_{+0.27}$ & $3.46^{-2.29}_{+0.60}$ & 
$-1.56^{-0.24}_{+0.24}$ & $0.96^{-0.09}_{+0.04}$ \nonumber\\ 
$a_{ij}^3$ & $1.81^{+0.52}_{-0.57}$ & $1.83^{+0.16}_{-0.16}$ & $-0.95^{+0.03}_{-0.03}$ & $1.10^{+0.78}_{-0.29}$ & $-2.56^{+2.30}_{-0.72}$ & 
$0.64^{+0.26}_{-0.24}$ & $0.03^{+0.09}_{-0.04}$  \nonumber\\ 
$b_{ij}$ & $-17.48^{-27.5}_{+4.25}$ & $2.20^{-2.15}_{+2.75}$ & $18.08^{-0.68}_{+0.66}$ & $42.18^{+1.31}_{-3.66}$ & $23.85^{+26.5}_{-0.78}$ & 
$36.27^{+4.72}_{-5.88}$ & $0.08^{-0.17}_{+2.84}$ \nonumber\\ 
\toprule
$b^{\bf ED}_1$ &  $0$ & & $39.24^{-2.10}_{+1.86}$ & & $39.24^{-2.10}_{+1.86}$ & & $39.24^{-2.10}_{+1.86}$  \nonumber\\
$b^{\bf ED}_2$ &  $0$ & & $18.85^{-1.13}_{+1.12}$ & & $18.85^{-1.13}_{+1.12}$ & & $18.85^{-1.13}_{+1.12}$ \nonumber\\
\hline
\end{tabular}}
\end{center}
\label{tab:3}
\end{table*}

With setting $q^2_{{\rm LCSR, max}} = 0.4 \,{\rm GeV}^2$, the fit result of SSE parameters are shown in table \ref{tab:3}. 
We mark that the superscript and subscript numbers are not the errors,  
but the differences to the central value fitted by the upper and lower predictions of the helicity form factors from LCSRs, respectively. 
We depict the modified helicity form factors in figure \ref{fig:2}, 
where the result obtained directly from LCSRs calculation is shown by lightblue bands, 
and the extrapolation by $z$-series parameterisation is shown by red bands. 
The form factors at end-point are obtained as ${\cal H}^{\rm SSE}_{\bf 00}(q_0^2) = 0$ and 
$\vert {\cal H}^{\rm SSE}_{ij \neq {\bf 00}}(q_0^2) \vert = 0.23^{+0.18}_{-0.23}$. 
The end-point constraints play an important role to set down the shapes of helicity form factors in the small recoiled regions 
where the LCSRs calculation is failed, in coordination with the kinematical structures in Eqs. (\ref{eq:helicityff-L}-\ref{eq:helicityff-T}). 
Besides the SSE parameterisation of the form factors, we have also tested the Becirevic and Kaidalov (BK) parameterisation \cite{Becirevic:1999kt} 
and found almost the same fit result of the helicity form factors. 

\begin{figure*}[h]
\begin{center}
\vspace{6mm}
\resizebox{0.4\textwidth}{!}{
\includegraphics{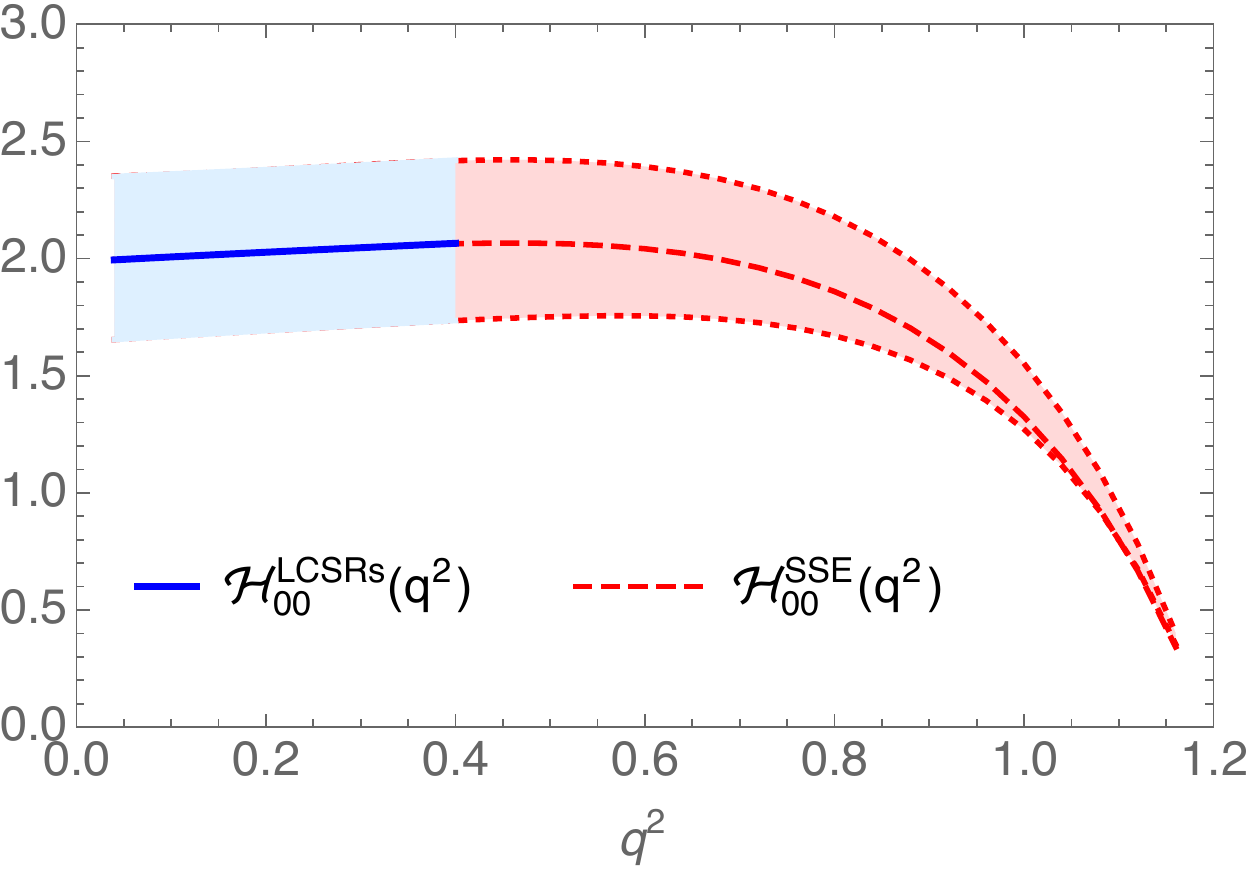}} 
\hspace{8mm}
\resizebox{0.4\textwidth}{!}{
\includegraphics{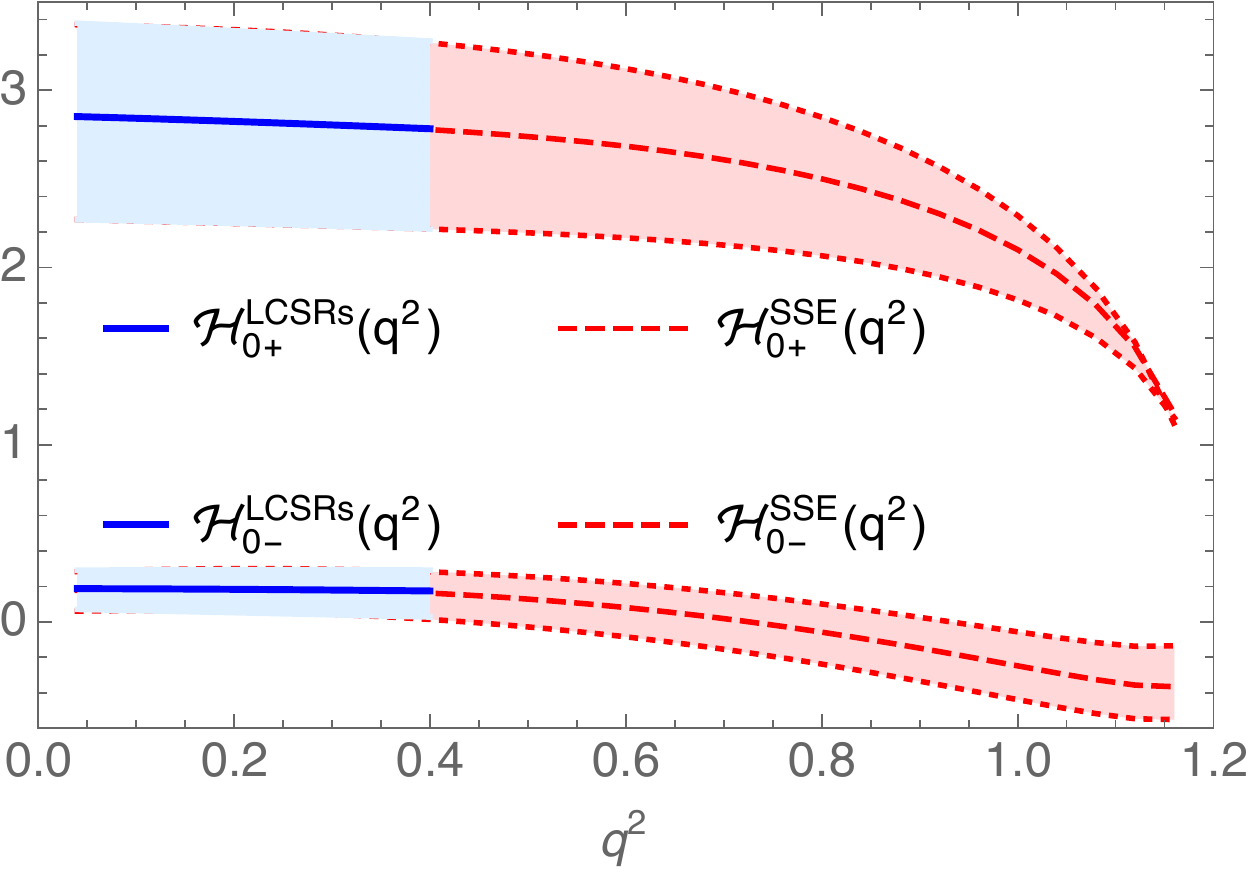}} \non
\vspace{2mm}
\resizebox{0.4\textwidth}{!}{
\includegraphics{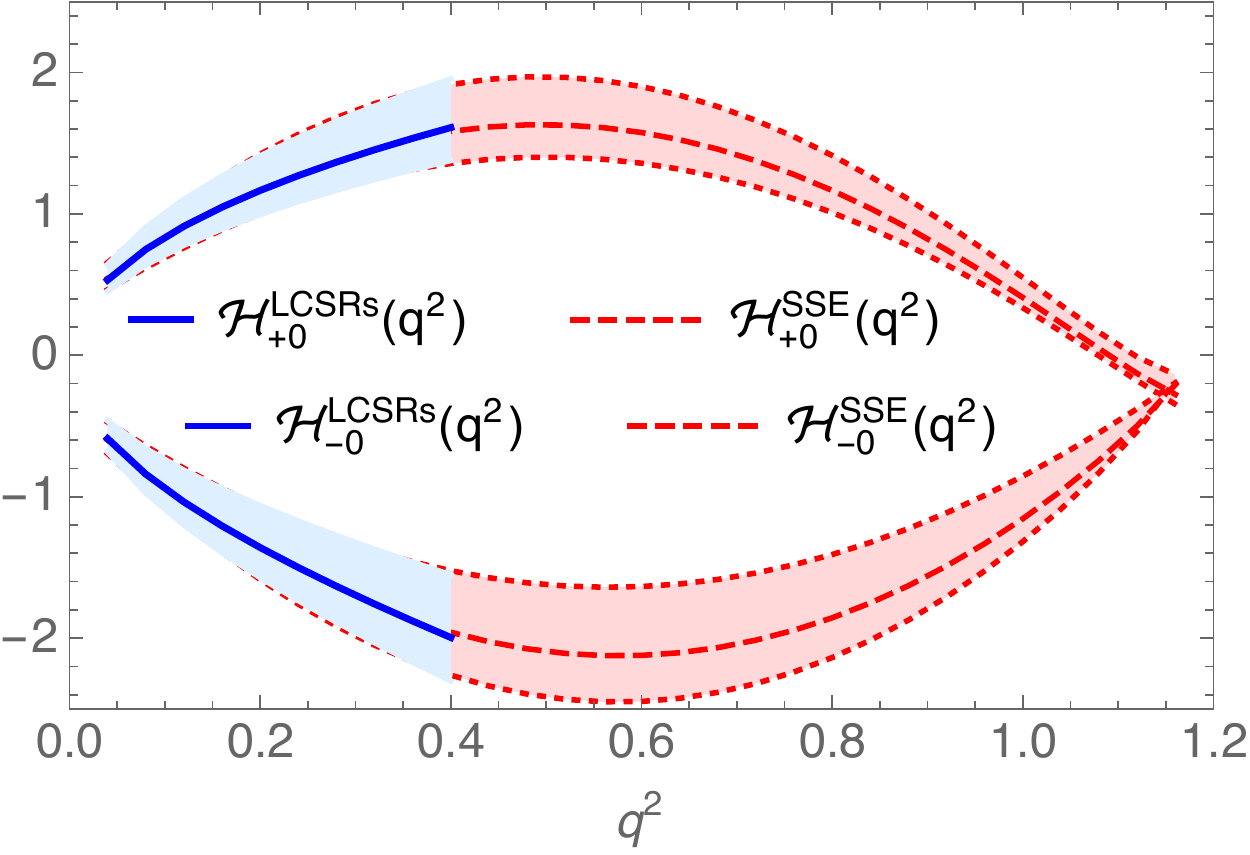}} 
\hspace{8mm}
\resizebox{0.4\textwidth}{!}{
\includegraphics{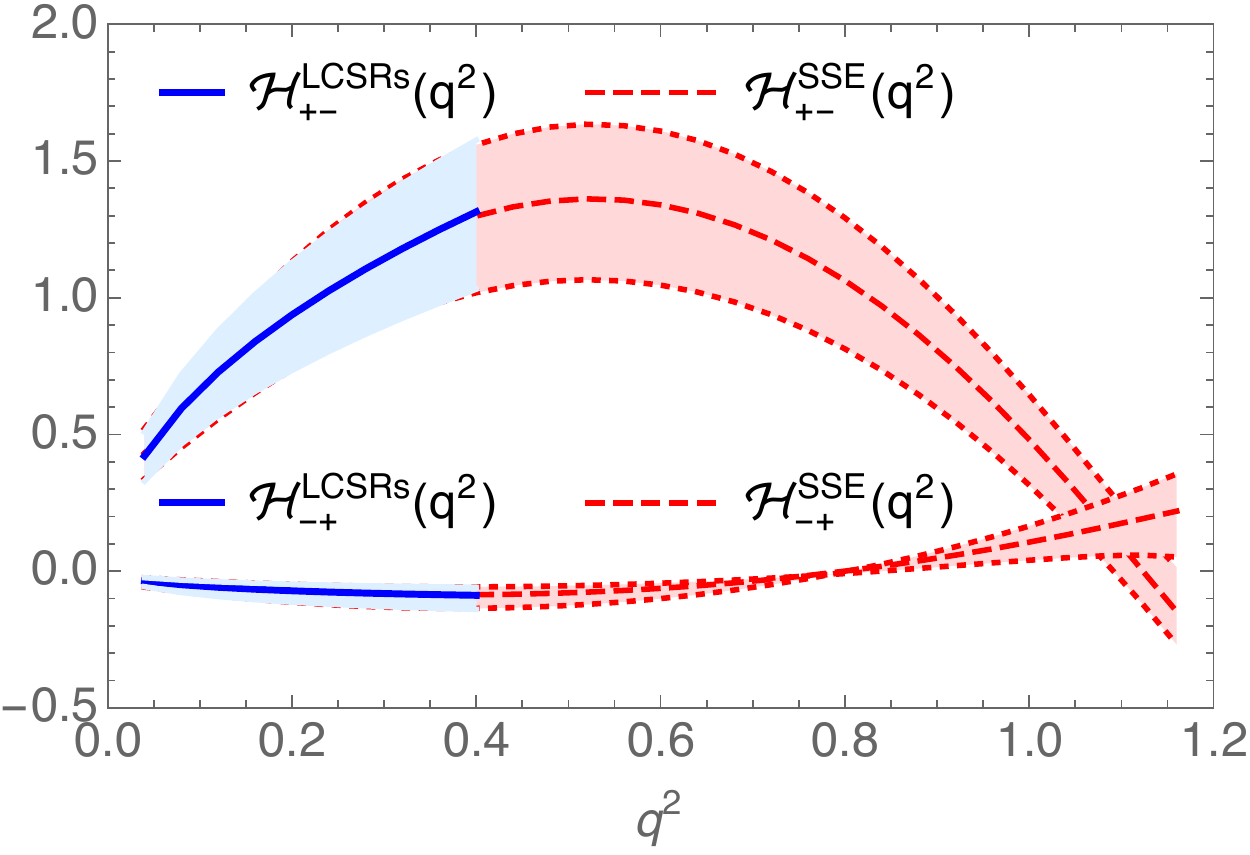}} \non
\end{center}
\vspace{-2mm}
\caption{The helicity form factors ${\mathcal H}_{ij}(q^2)$ obtained from the LCSRs calculation in the large recoiled regions 
and the extrapolating to the whole kinematical region by SSE parameterization.}
\label{fig:3}       
\end{figure*}

\section{Exclusive $D_s^\ast$ weak decays} 

The leptonic decays $D_s^\ast \to \ell \nu\ (\ell = e, \mu)$ have the decay width 
\begin{eqnarray}
\Gamma_{D_s^\ast \to l \nu} &=& \frac{G_F^2}{12 \pi} \vert V_{cs} \vert^2 f_{D_s^\ast}^2 m_{D_s^\ast}^3 
= 2.44 \times 10^{-12} \; {\rm GeV} \; ,
\label{eq:Dss2lnu}
\end{eqnarray}
if we accept the lattice result of the decay constant $f_{D_s^\ast} = 0.274 \, {\rm GeV}$ \cite{Donald:2013sra} and neglect the lepton masses.
The differential decay width of semileptonic decays of a particular polarization mode is written as
\begin{eqnarray}
&&\frac{d\Gamma_{ij}}{d q^2} = \frac{G^2_F \vert V_{cs} \vert^2 \lambda^{1/2} q^2 }{192 \pi^3 m_{D_s^\ast}^3} 
\vert {\cal H}_{ij}(q^2)\vert^2 \; .
\label{eq:widths}
\end{eqnarray}
With the helicity form factors obtained above, we obtain the spin averaged total decay width
\begin{eqnarray}
\Gamma_{D_s^\ast \to \phi l \nu_l} &=& \frac{1}{3} \int_{0}^{q_0^2} d q^2 \sum_{i,j=0,\pm} \frac{d \Gamma_{ij}}{d q^2}  \nonumber\\
&=& (3.28^{+0.82}_{-0.71}) \times 10^{-14} \, {\rm GeV}. 
\label{eq:Dss2philnu-width}
\end{eqnarray}
The leptonic and semileptonic $D_s^\ast$ weak decays, meanwhile, extent the investigation of lepton flavour university (LFU) study 
\cite{HFLAV:2019otj,BESIII:2019vhn,BESIII:2018nzb}. 

Under the naive factorisation hypothesis with considering only the color singlet operator at tree level, 
the decay amplitudes of $D_s^\ast \to \phi \pi, \phi \rho$ channels are detached into two matrix elements, 
\begin{eqnarray}
&&{\mathcal A}_{D_s^{\ast +} \to \phi \pi^+} = (-i) \frac{G_F}{\sqrt{2}} V_{cs} a_1 m_\pi f_\pi \sum_{j=0,\pm} \mathcal{H}_{0j}(m_\pi^2), \nonumber\\
&&{\mathcal A}_{D_s^{\ast +} \to \phi \rho^+} = \frac{G_F}{\sqrt{2}} V_{cs} a_1 m_\rho f^{\parallel}_\rho \sum_{i,j=0,\pm} \mathcal{H}_{ij}(m_\rho^2).
\label{eq:Dss2phirho(pi)_amp}
\end{eqnarray}
Considering the wilson coefficient $a_1 = 0.999$ at the factorisation scale $\mu = (m_{D_s^\ast}^2-m_c^2)^{1/2}$ \cite{Buchalla:1995vs} 
and the decay constants $f_\pi = 0.130 \, {\rm GeV}$ \cite{PDG2022} and $f_\rho^{\parallel}= 0.210 \, {\rm GeV}$ \cite{Bharucha:2015bzk}, 
we obtain the partial widths of nonleptonic decays as 
\begin{eqnarray}
&&\Gamma_{D_s^{\ast +} \to \phi \pi^+} = ( 3.81^{+1.52}_{-1.33} ) \times 10^{-14} \, {\rm GeV},   \nonumber\\
&&\Gamma_{D_s^{\ast +}\to \phi \rho^+} = ( 1.16^{+0.42}_{-0.39} ) \times 10^{-13} \, {\rm GeV}.
\label{eq:width_Dss2phirho}
\end{eqnarray}
The large uncertainty ($40 \%$) in $\phi\pi$ channel comes from the LCSRs predictions of the helicity form factors at the momentum transfer point $q^2 = m_\pi^2$, 
meanwhile the uncertainty ($\sim 40\%$) in $\phi\rho$ channel comes from the extrapolation by simplified $z$-series expansion at the momentum transfer point $q^2 = m_\rho^2$. 
The prediction of $\Gamma_{D_s^{\ast +} \to \phi \pi^+}$ is marginally consistent with 
the recent calculation based on the perturbative QCD approach \cite{Yang:2022esh}, but is half smaller in the magnitude. 
We mark that the color mixing operator ($a_2$ proportional) at tree level and the non-perturbative contributions are usually significant, 
and could give sizable contributions to the hadronic decays, accompanying with the contribution from timelike polarisation of leptonic current ${\bar \epsilon}_\mu(t)$. 
We postpone these contributions for the further study.
If we take the total width $\Gamma_{D_s^\ast} = (7.0 \pm 2.8) \times 10^{-8} \, {\rm GeV}$ evaluated from lattice QCD \cite{Donald:2013sra}, 
the branching fractions of $D_s^\ast$ weak decays are
\begin{eqnarray}
&&{\cal B}(D_s^\ast \to l \nu) = (3.49 \pm 0.14) \times 10^{-5},  \nonumber\\
&&{\cal B}(D_s^\ast \to \phi l \nu) = (0.47^{+0.12}_{-0.10} \pm 0.19) \times 10^{-6},  \nonumber\\
&&{\cal B}(D_s^\ast \to \phi \pi) = (0.54^{+0.22}_{-0.19} \pm 0.22) \times 10^{-6},  \nonumber\\
&&{\cal B}(D_s^\ast \to \phi \rho) = (1.65^{+0.61}_{-0.56} \pm 0.66) \times 10^{-6}. 
\end{eqnarray}

Let us give a brief discussion on the experimental potential of $D_s^\ast$ weak decays. 
The integrated luminosity at Belle II would achieve $10 \, {\rm ab}^{-1}$ after the phase 3 running (2024-2026) \cite{Belle-II:2018jsg},
which would produce an available $D_s(D_s^\ast)$ sample at order ${\cal O}(10^9)$ 
by considering ${\cal O}(10^6)$ $D_s \to \phi(KK) \pi^+$ signals (with efficiency $22\%$) 
are obtained based on $921 \, {\rm fb}^{-1}$ data sample \cite{Dss_sample,Belle:2021ygw} 
and the branching fraction $B(D_s \to \phi(KK) \pi) = 2.24 \%$ \cite{PDG2022}.
With this sample, about ${\cal O}(10^7)$ (${\cal O}(10^2)$) signals of $D_s (D_s^\ast) \to \phi(KK) \pi$ would be obtained,
which indicates the feasibility of searching for $D_s^\ast \to \phi \pi$ at Belle II.
Meanwhile, about $3.07 \times 10^{6}$ $D_s^\ast$ mesons have been collected by BESIII 
with the integrated luminosity $3.2 \, {\rm fb}^{-1}$ at $4.178 \, {\rm GeV}$ \cite{BESIII:2020nme}.
They are directly produced from the $e^+e^-$ collision at the $D_sD_s^\ast$ threshold with lower background, 
and it provides a good chance to measure the leptonic decays $D_s^\ast \to l \nu$ and to further determine $\Gamma_{D_s^\ast}$. 
Note that the photon-radiation effect is tiny in the leptonic $D_s^\ast$ decays 
since these channels are not helicity suppressed in contrast to the $D_s \to l \nu$ decays. 
For the hadronic decay channels, 
we hope LHCb, with the excellent particle identification to distinguish $K, \pi$ and $\mu$, 
would study the $D_s^\ast \to \phi(KK) \pi$ channel with $D_s^\ast$ producing from semileptonic decay 
$B_s \to D_s^\ast \mu \nu$ \cite{LHCb:2020hpv}. 

\section{Summary}

In this work we calculate the $D_s^\ast \to \phi$ helicity form factors from LCSRs 
with the accuracy up to two-particle twist-5 DAs of the $\phi$ meson at the leading order of $\alpha_s$, 
with which we study the experimental potential of discovering $D_s^\ast$ weak decays. 
The result shows that the leptonic decays $D_s^\ast \to l \nu$ are the most hopeful channels to be measured at BESIII, 
the semileptonic decays $D_s^\ast \to \phi l \nu$ could be accessible at Belle II after the phase 3 running, 
and the hadronic $D_s^\ast \to \phi\pi$ decays are promising at LHCb.
The measurement of purely leptonic decays would determine the total width of the $D_s^\ast$ meson 
and hence clarify some fundamental properties of the $D_s^\ast$ meson, 
such as the electromagnetic and strong couplings $g_{D_s^\ast D_s \gamma}$ and $g_{D_s^\ast D_s \pi}$. 
It is highly hopeful that these channels will promote the first observation of weak decays of a vector meson, 
opening up a new playground to test the standard model and pushing us to higher precision studies.

\section*{Acknowledgments}
We are grateful to Qin Chang, Long-ke Li, Hai-long Ma, Wei Shan, Chen-ping Shen and Lei Zhang for fruitful discussions, 
especially to Liang Sun for proposing the $D_s^\ast \to \phi(KK) \pi$ channel at LHCb. 
This work is partly supported by the National Natural Science Foundation of China (NSFC) under Grant 
Nos. 11805060, 11975112, 12005068, and the Joint Large Scale Scientific Facility Funds of the NSFC and CAS under Contract No. U1932110. 
S.C. is also supported by the Natural Science Foundation of Hunan Province, China under Grant No. 2020JJ4160.

\begin{appendix}

\section{Definition of $\phi$ meson on the light cone}\label{LCDAs-phi}

In order to facilitate the light-cone expansion, the meson four-momentum ($p_\mu$) and close to lightlike separation ($x_\mu$) 
are expressed as linear combinations of the lightlike vectors (${\hat p}_\mu, z_\mu$) \cite{Chernyak:1983ej}, 
\beq
&&{\hat p}_\mu = p_\mu - \frac{1}{2} z_\mu \frac{m_M^2}{{\hat p} \cdot z}  \,, \nonumber\\
&&z_\mu 
= x_\mu \left[ 1 - \frac{x^2 m_M^2}{4({\hat p} \cdot z)^2} \right] - \frac{1}{2} {\hat p}_\mu \frac{x^2}{{\hat p} \cdot z} + \mathcal{O}(x^4)\,.
\label{eq:lc-expansion}
\eeq
Meanwhile, the polarization vectors decompose into three terms, 
\beq
\epsilon_\mu^\lambda &=& \frac{\epsilon^\lambda \cdot x}{{\hat p} \cdot z} \, {\hat p}_\mu + \frac{\epsilon^\lambda \cdot {\hat p}}{{\hat p} \cdot z }\, z_\mu + \epsilon^\lambda_{\perp \mu} \nonumber\\
&=& (\epsilon^\lambda \cdot x) \, \frac{p_\mu (p \cdot x) - x_\mu m_M^2}{(p \cdot x)^2 - x^2 m_M^2} \,.  
\label{eq:pola}
\eeq
Here $\lambda$ is the polarisation of meson and we use the symbols $\parallel, \perp$ 
to denote the longitudinal and transversal directions of the polarizations, respectively. 

LCDAs are rigorously defined by the matrix element sandwiched with the quark bilinears with light-cone separation, 
and then switch to the actual momenta and the near lightlike distance $x$ for the practise of phenomenas. 
In Refs. \cite{Bharucha:2015bzk,Ball:1998sk,Ball:1998ff}, 
high twist LCDAs of vector mesons are systematical studied in QCD 
based on conformal expansion with taking into account meson and quark mass corrections.
The complete analysis of the parameters from QCD sum rules and a renormalon based model are presented in Refs. \cite{Ball:2007rt,Ball:2007zt}. 
Considering the polarisation decomposition in Eq. (\ref{eq:pola}) to the vector Dirac structure, 
the matrix element sandwiched between vacuum and vector meson state takes the following parameterisation
\beq
&~&\langle \phi(p,\epsilon) \vert {\bar s}(x) \gamma_\mu s(0) \vert 0 \rangle \nonumber\\
&=& f_\phi^\parallel m_\phi \int_0^1 du e^{i u p \cdot x} \Big\{ 
\epsilon_\mu \left[ \phi_3^\perp(u) + \frac{m_\phi^2 x^2}{16} \phi_5^\perp(u) \right]  \nonumber\\
&+& p_\mu \frac{\epsilon \cdot x}{p \cdot x} 
\left[ \phi_2^\parallel(u) - \phi_3^\perp(u) + \frac{m_\phi^2 x^2}{16}  \left( \phi_4^\parallel(u) - \phi_5^\perp(u) \right) \right]  \nonumber\\
&-& \frac{(\epsilon \cdot x) m_\phi^2}{2 (p \cdot x)^2} x_\mu \Big[ \psi_4^\parallel(u) - 2 \phi_3^\perp(u) + \phi_2^\parallel(u) \Big] \Big\} \,, 
\label{eq:LCDAs-definition-1}\\
&~&\langle \phi(p,\epsilon) \vert {\bar s}(x) \sigma_{\mu\nu} s(0) \vert 0 \rangle \nonumber\\
&=& - i f^\perp_\phi \int_0^1 du e^{i u p \cdot x} \Big\{ 
\left( \epsilon_\mu p_\nu - \epsilon_\nu p_\mu \right) \left[ \phi_2^\perp(u) + \frac{m_\phi^2 x^2}{16} \phi_4^\perp(u) \right]  \nonumber\\
&+& \left( p_\mu x_\nu - p_\nu x_\mu \right) \frac{(\epsilon \cdot x) m_\phi^2}{(p \cdot x)^2} 
\left[ \phi_3^\parallel(u) - \frac{1}{2} \phi_2^\perp(u) - \frac{1}{2} \psi_4^\perp(u)\right]  \nonumber\\
&+& \frac{m_\phi^2}{2(p \cdot x)}\left( \epsilon_\mu x_\nu - \epsilon_\nu x_\mu \right) \Big[ \psi_4^\perp(u) - \phi_2^\perp(u) \Big] \Big\} \,, 
\label{eq:LCDAs-definition-2}\\
&~&\langle \phi(p,\epsilon) \vert {\bar s}(x) \gamma_\mu \gamma_5 s(0) \vert 0 \rangle \nonumber\\
&=& - \frac{ f_\phi^\parallel m_\phi \varepsilon_{\mu\nu\rho\sigma} \epsilon^{\ast \nu} p^\rho x^\sigma}{4} \int_0^1du e^{i u p \cdot x} \nonumber\\
&~& \cdot \left[ \tilde{\psi}_3^\perp(u) + \frac{m_\phi^2 x^2}{16} \tilde{\psi}_5^\perp(u) \right] \,,
\label{eq:LCDAs-definition-2}\\
&~&\langle \phi(p,\epsilon) \vert {\bar s}(x) s(0) \vert 0 \rangle = - \frac{i}{2} f_\phi^\perp (\epsilon \cdot x) m_\phi^2 \int_0^1du e^{i u p \cdot x} \tilde{\psi}_3^\parallel(u) \,.
\eeq
We use the conventions 
\beq
\varepsilon_{0123} = - \varepsilon^{0123} = 1, \quad
\gamma_5 = \frac{i}{4!} \varepsilon^{\mu\nu\rho\sigma} \gamma_\mu \gamma_\nu \gamma_\rho \gamma_\sigma.
\eeq
The quark mass effects are taken into account in the last two parameterisations of matrix elements 
with the Dirac structures $\gamma_\mu \gamma_5$ and ${\bf 1}$, and the auxillary DAs read as 
\beq 
&&\tilde{\psi}_3^\parallel(u) = \left( 1 - r_\parallel \delta_+ \right) \psi_3^\parallel(u)\,, \nonumber\\
&&\tilde{\psi}_{3(5)}^\perp(u) = \left( 1 - r_\perp \delta_+ \right) \psi_{3(5)}^\perp(u)\,, 
\eeq
with $r_\parallel = f_\phi^\parallel/f_\phi^\perp$, $r_\perp = f_\phi^\perp/f_\phi^\parallel$ and $\delta_\pm = \left( m_s \pm m_s \right)/m_\phi$. 
The DAs $\phi = \{ \phi_2^{\parallel(\perp)}, \phi_3^{\parallel(\perp)}, \psi_3^{\parallel(\perp)}, \psi_4^{\parallel(\perp)} \}$ 
satisfy the normalisations 
\beq
\int_0^u \phi(u^\prime) du^\prime \Big\vert_{u=1} = 1, \quad  
\int_0^u du^\prime \int_0^{u^\prime} \phi(u^{\prime\prime}) du^{\prime\prime} \Big\vert_{u=1} = 1 \,
\eeq 
and also the equation of motions (EOM) of the LCDAs. 
The DAs $\phi^\prime = \{ \phi_4^{\parallel(\perp)}, \phi_5^\perp, \psi_5^\perp \}$ 
are not subject to a particular normalisation while $\int_0^u du^\prime \left( \phi_4^\parallel - \phi_5^\perp \right) \big\vert_{u=1} = 0 $ is necessary, 
and they relates to DAs $\phi$, at first order of ${\cal O}(m_\phi^2)$ expansion, by \cite{Ball:1998ff}
\beq
\phi_4^\parallel(u) &=& - 4 \int_0^u \left[ (2 u^\prime -1) \phi_2^\parallel(u^\prime) \right] \nonumber\\ 
&+& 4 \int_0^u du^\prime \int_0^{u^\prime} du^{\prime\prime} 
\left[ \phi_3^\perp(u^{\prime\prime}) - \psi_4^\parallel(u^{\prime\prime}) - 3 \phi_2^\parallel(u^{\prime\prime}) \right] \,,  \nonumber\\
\phi_4^\perp(u) &=& - 4 \int_0^u \left[ (2 u^\prime -1) \phi_2^\perp(u^\prime) \right] \nonumber\\ 
&+& 4 \int_0^u du^\prime \int_0^{u^\prime} du^{\prime\prime} \left[ \psi_4^\perp(u^{\prime\prime}) - \phi_2^\perp(u^{\prime\prime}) \right] \,,  \nonumber\\
\phi_5^\perp(u) &=& - 4 \int_0^u \left[ (2 u^\prime -1) \phi_3^\perp(u^\prime) \right] \,, \nonumber\\ 
\psi_5^\perp(u) &=& - 4 \int_0^u \left[ (2 u^\prime -1) \psi_3^\perp(u^\prime) \right] \,.
\label{eq:phi-phip-relations}
\eeq
We notice that the last two relation equations hold only for asymptotic LCDAs \cite{Bharucha:2015bzk}.

\begin{table*}[t]
\caption{Notations of the LCDAs of light vector mesons (up tabular) and 
nonperturbative parameters at the factorization scale $\mu_f = 1.66 \, {\rm GeV}$ of the $\phi$ meson LCDAs taken in our evaluation (low tabular).}
\begin{center}
\setlength{\tabcolsep}{2mm}{
\begin{tabular}{l | c c | c c c c | c c c c | c c}
\hline
${\rm Notations}$ & $\phi_2^\parallel$ & $\phi_2^\perp$ & $\phi_3^\parallel$ & $\psi_3^\parallel$ & $\phi_3^\perp$ & $\psi_3^\perp$ & 
$\phi_4^\parallel$ & $\psi_4^\parallel$ & $\phi_4^\perp$& $\psi_4^\perp$ & $\phi_5^\perp$ & $\psi_5^\perp$  \nonumber\\ 
\hline
${\rm Twist}$ & 2 & 2 & 3 & 3 & 3 & 3 & 4 & 4 & 4 & 4 & 5 & 5  \nonumber\\ 
\hline
${\rm Dirac}$ & $\gamma_\mu$ & $\sigma_{\mu\nu}$ & $\sigma_{\mu\nu}$ & ${\bf 1}$ & $\gamma_\mu$ & $\gamma_\mu \gamma_5$ & 
$\gamma_\mu$ & $\gamma_\mu$ & $\sigma_{\mu\nu}$ & $\sigma_{\mu\nu}$ & $\gamma_\mu $ & $\gamma_\mu \gamma_5$  \nonumber\\
\hline
${\rm Expressions}$ & {\rm (\ref{eq:phi2})} & {\rm (\ref{eq:phi2})} & {\rm (\ref{eq:phi3para})} & {\rm (\ref{eq:psi3para})} & {\rm (\ref{eq:phi3perp})} & {\rm (\ref{eq:psi3perp})} & {\rm (\ref{eq:phi4})} & {\rm (\ref{eq:psi4})} & {\rm (\ref{eq:phi4})} & {\rm (\ref{eq:psi4})} & {\rm (\ref{eq:phi5})} & {\rm (\ref{eq:phi5})}  \nonumber\\ 
\hline
\end{tabular}\\
\vspace{6mm}
\begin{tabular}{l | c c | c c | c c c}
\hline
${\rm Para.}$  & ${\bar m}_s({\rm GeV})$  &  $m_\phi({\rm GeV})$ & $f_\phi^\parallel({\rm GeV})$ & $f_\phi^\perp({\rm GeV})$  & 
$a_1^{\parallel(\perp)}$  & $a_2^\parallel$ & $a_2^\perp$  \nonumber\\ 
\hline
${\rm Value}$ & $0.101(8)$ & $1.68$ & $0.233(4)$ & $0.184(4)$ & $0$ & $0.243(80) $ & $0.148(70)$  \nonumber\\
\hline
\end{tabular}}
\end{center}
\label{tab:4}
\end{table*}

The lower twists DAs are conventionally expanded in conformal spin which is analogous to the partial wave expansion of ${\rm SO(3)}$, 
and write in terms of Gegenbauer polynomials with corresponding moments. 
In this work we take the truncation to the second order of leading twist DAs expansion, 
\beq
\phi_2^{\parallel(\perp)}(u) = 6u(1-u) \left[ 1 + a_1^{\parallel(\perp)} C_1^{3/2}(t) + a_2^{\parallel(\perp)} C_2^{3/2}(t) \right]\,.
\label{eq:phi2}
\eeq
The twist 3 DAs contributed from the leading twist DAs are cited as \cite{Ball:1998sk}
\beq
&&\phi_3^\parallel(u) = \frac{1}{2} \int_0^u du^\prime \frac{\Psi^\parallel_2(u^\prime)}{{\bar u}^\prime} 
+ \frac{1}{2} \int_u^1 du^\prime \frac{\Psi^\parallel_2(u^\prime)}{u^\prime}  \,, \label{eq:phi3para} \nonumber\\
&&\tilde{\psi}_3^\parallel(u) = {\bar u} \int_0^u du^\prime \frac{\Psi^\parallel_2(u^\prime)}{{\bar u}^\prime} 
+ u \int_u^1 du^\prime \frac{\Psi^\parallel_2(u^\prime)}{u^\prime}  \,, \label{eq:psi3para} \nonumber\\
&&\phi_3^\perp(u) = \frac{1}{4} \int_0^u du^\prime \frac{\Psi^\perp_2(u^\prime)}{{\bar u}^\prime} 
+ \frac{1}{4} \int_u^1 du^\prime \frac{\Psi^\perp_2(u^\prime)}{u^\prime} \,, \label{eq:phi3perp} \nonumber\\
&&\tilde{\psi}_3^\perp(u) = {\bar u} \int_0^u du^\prime \frac{\Psi^\perp_2(u^\prime)}{{\bar u}^\prime} 
+ u \int_u^1 du^\prime \frac{\Psi^\perp_2(u^\prime)}{u^\prime}  \,, \label{eq:psi3perp}
\eeq
with the auxiliary functions 
\beq
&&\Psi^\parallel_2(u^\prime) = 2 \phi_2^\perp(u^\prime)+ r_\parallel \left[ \frac{\left(3-2 u^\prime\right)}{2} \delta_+  + \frac{\delta_-}{2} \right] 
\frac{\partial \phi_2^\perp(u^\prime)}{\partial u^\prime} \,,  \nonumber\\
&&\Psi^\perp_2(u^\prime) = 2 \phi_2^\parallel(u^\prime) + r_\perp \left[ \delta_+ (2 u^\prime -1) + \delta_- \right] \frac{\partial \phi_2^\perp(u^\prime)}{\partial u^\prime} \,.
\label{eq:psi2to3}
\eeq
We here present their explicit expressions truncated to the second term of gegenbauer polynormias of $\phi_2^{\parallel/\perp}(u)$
\beq
\phi_3^\parallel(u) &=& 3-6 u+6 u^2 \nonumber\\ 
&+& \frac{3 r_\parallel}{2} \delta_+ \left[-6+12 u-4 u^2+\log(1-u)-3 \log u \right]  \nonumber\\
&+& a_2^\perp(\mu) \left[ 3 (1-12 u+42 u^2-60 u^3+30 u^4) \right]  \nonumber\\
&+& a_2^\perp(\mu) 3 r_\parallel \delta_+ \left[ -33+126 u-222 u^2+200 u^3 \right. \nonumber\\ 
&~& \left. -60 u^4+3 \log(1-u)-9 \log u \right] \,,
\label{eq:phi3para-1} \\
\tilde{\psi}_3^\parallel(u) &=& 2 (1-u) \left[ 3-6 u+6 u^2 \right. \nonumber\\ 
&~& \left. + \frac{3 r_\parallel}{2} \delta_+ (-6+12 u-4 u^2+ \log(1-u) - 3 \log u ) \right]  \nonumber\\
&+& a_2^\perp (\mu) 6 u \left[ 1-12 u+42 u^2-60 u^3+30 u^4 \right]  \nonumber\\
&+& a_2^\perp(\mu) 6 u r_\parallel \delta_+ \left[ -33+126 u-222 u^2+200 u^3 \right. \nonumber\\ 
&~& \left. -60 u^4+3 \log(1-u)-9\log u \right] \,,
\label{eq:psi3para-1} 
\eeq
\beq
\phi_3^\perp(u) &=&  \frac{1}{2} \Big\{ 3-6 u+6 u^2 \nonumber\\ 
&+& 3 r_\perp \delta_+ \left[ 2-4 u+4 u^2+ \log(1-u)+ \log u \right] \nonumber\\ 
&+& a_2^\parallel(\mu) 3 \left[ 1-12 u+42 u^2-60 u^3+30 u^4 \right]  \nonumber\\
&+& a_2^\perp(\mu) 6 r_\perp \delta_+ \left[ 11-42 u+102 u^2-120 u^3 \right. \nonumber\\ 
&~& \left. +60 u^4+3 \log(1-u) + 3 \log u \right] \Big\} \,,
\label{eq:phi3perp-1} \\
\tilde{\psi}_3^\perp(u) &=& 2 (1-u) \left[ 3-6 u+6 u^2 \right. \nonumber\\ 
&~& \left. + 3 r_\perp \delta_+ (2-4u+4u^2+\log(1-u)+\log u ) \right]  \nonumber\\
&+& a_2^\parallel(\mu) 6 u \left[ 1-12 u+42 u^2-60 u^3+30u^4 \right] \nonumber\\
&+& a_2^\perp(\mu) 12 u  r_\perp \delta_+ \left[ 11-42 u+102 u^2-120 u^3 \right. \nonumber\\ 
&~& \left. +60 u^4+3 \log(1-u) + 3 \log u \right] \,.
\label{eq:psi3perp-1}
\eeq

In the asymptotic limit, the twist 4 DAs contributed at the order ${\cal O}((p \cdot x)^{-2})$ are given by 
\beq
\psi_4^\parallel(u) = 6 u {\bar u} \,, \quad\quad\quad\quad\quad\, \psi_4^\perp(u) = 6 u {\bar u} \,,
\label{eq:psi4}
\eeq
the twist 4 and twist 5 DAs $\phi^\prime$ contributed at the order ${\cal O}(m_\phi^2x^2)$ read from Eq. (\ref{eq:phi-phip-relations}) as 
\beq
&&\phi_4^\parallel(u) = 24 u^2 {\bar u}^2 \,, \quad \phi_4^\perp(u) = 24 u^2 {\bar u}^2 \,,  \label{eq:phi4} \nonumber\\
&&\phi_5^\perp(u) = 6 u {\bar u} \left( 1 - u {\bar u} \right) \,, \quad \psi_5^\perp(u) = 12 u^2 {\bar u}^2 \,. \label{eq:phi5}
\eeq

For the sake of convenient we list the DAs at different twists with corresponding Dirac structures in table \ref{tab:4}, 
and also we present the nonperturbaive parameters of LCDAs taken in our evaluation. 
The mass of $\phi$ meson and strange quark in the ${\rm \overline{ MS}}$ scheme are taken from PDG \cite{PDG2022}. 
The longitudinal decay constant $f_\phi^\parallel$ is mainly determined directly by the experiment measurement of channel 
$e^+e^- \to \phi (\to PP)$ \cite{Bharucha:2015bzk}, 
the scale dependent transversal decay constant is chosen by considering the ratio 
$r_\perp(1 \, {\rm GeV}) = f_\phi^\perp (1 \, {\rm GeV})/f_\phi^\parallel = 0.820$ obtained from lattice QCD 
simulated by using $N_f = 2 + 1$ domain-wall fermions at the spacing $a = 0.114 \, {\rm fm}$ and masses down
to $m_\pi = 330 \, {\rm MeV}$ \cite{RBC-UKQCD:2008mhs}. 
The Gegenbauer moments $a_2^{\parallel(\perp)}$ are taken from Ref. \cite{Dimou:2012un} where a
combined analysis is performed on the lattice simulation and QCD sum rule calculation \cite{Arthur:2010xf}.

\section{Imaginary part of OPE invariant amplitudes}\label{app:imaginary-amplitudes}

In the case of transversal helicity with longitudinal leptonic current and transversal $\phi$ meson, 
the imaginary part of OPE invariant amplitudes are
\beq
&~&\frac{1}{\pi} \, {\rm Im} F^{{\rm OPE}}_{1,{\bf 0 \pm}}(q^2<0, u)  \nonumber\\
&=& \left[ \frac{\sqrt{\lambda}}{2 \sqrt{\vert q^2 \vert}} \pm \frac{\left( m_{D_s^\ast}^2 - m_\phi^2 -q^2 \right)}{2 \sqrt{\vert q^2 \vert}} \right] 
m_c f_\phi^\perp \, \phi_2^\perp(u) \nonumber\\
&+& \left[ \frac{u\sqrt{\lambda}}{2 \sqrt{\vert q^2 \vert}} \pm \frac{u \left( m_{D_s^\ast}^2 - m_\phi^2 -q^2 \right) + 2 q^2}{2 \sqrt{\vert q^2 \vert}} \right] 
f_\phi^\parallel m_\phi \, \phi_3^\perp(u) \nonumber\\ 
&+& \left[ - \frac{\sqrt{\lambda}}{2 \sqrt{\vert q^2 \vert}} \pm \frac{\left( m_{D_s^\ast}^2 - m_\phi^2 -q^2 \right)}{2 \sqrt{\vert q^2 \vert}} \right] 
f_\phi^\parallel m_\phi \left( {\bar \phi}_2^\parallel(u) - {\bar \phi}_3^\perp(u) \right) \nonumber\\
&+& \frac{\sqrt{\lambda}}{4 \sqrt{\vert q^2 \vert}} \, f_\phi^\parallel m_\phi \, \tilde{\psi}_3^\perp(u)  \,,
\label{eq:Im-OPE-LT-n1} \\
&~&\frac{1}{\pi} \,{\rm Im} F^{{\rm OPE}}_{2,{\bf 0 \pm}}(q^2<0, u) \nonumber\\
&=& \left[ \frac{\sqrt{\lambda} \left[ u \left(m_{D_s^\ast}^2 - m_\phi^2 - q^2 \right) + 2 q^2 \right]}{2 \sqrt{\vert q^2 \vert}} \pm \frac{u \lambda}{2 \sqrt{\vert q^2 \vert}} \right] \nonumber\\
&~& \cdot \Big[ \frac{f_\phi^\parallel m_\phi \, \tilde{\psi}_3^\perp(u)}{4} \mp  \frac{f_\phi^\parallel m_\phi^3 \, \phi_5^\perp(u) }{4 \sqrt{\lambda}} \nonumber\\
&~& \; \mp \frac{m_c f_\phi^\perp m_\phi^2 \left({\bar \psi}_4^\perp(u) - {\bar \phi}_2^\perp(u) \right)}{\sqrt{\lambda}} \Big] \nonumber\\ 
&-& \frac{u \sqrt{\lambda}}{\sqrt{\vert q^2 \vert}} \, f_\phi^\parallel m_\phi^3 
\left( \overset{=}{\psi^\parallel_4}(u) +  \overset{=}{\phi^\parallel_2}(u) - 2 \overset{=}{\phi^\perp_3}(u) \right) \nonumber\\
&+& \left[ \frac{\sqrt{\lambda}}{8 \sqrt{\vert q^2 \vert}} \mp \frac{\left( m_{D_s^\ast}^2 - m_\phi^2 -q^2 \right)}{8 \sqrt{\vert q^2 \vert}} \right] 
f_\phi^\parallel m_\phi^3 \left( {\bar \phi}_4^\parallel(u) - {\bar \phi}_5^\perp(u) \right)  \nonumber\\ 
&-& \frac{\sqrt{\lambda} }{16 \sqrt{\vert q^2 \vert}} \, f_\phi^\parallel m_\phi^3 \, \tilde{\psi}_5^\perp(u) 
+ \left[ \frac{\sqrt{\lambda}}{2 \sqrt{\vert q^2 \vert}} \pm \frac{\left( m_{D_s^\ast}^2 - m_\phi^2 -q^2 \right)}{2 \sqrt{\vert q^2 \vert}} \right] \nonumber\\
&~& \cdot f_\phi^\perp m_\phi^2 m_c \left( \overset{=}{\psi^\perp_4}(u) +  \overset{=}{\phi^\perp_2}(u) - 2 \overset{=}{\phi^\parallel_3}(u) \right) \,,
\label{eq:Im-OPE-LT-n2} 
\eeq
\beq
&~&\frac{1}{\pi} \,{\rm Im} F^{{\rm OPE}}_{3,{\bf 0 \pm}}(q^2<0, u) \nonumber\\
&=& \left[ - \frac{\sqrt{\lambda}}{4 \sqrt{\vert q^2 \vert}} \mp \frac{\left(m_{D_s^\ast}^2 - m_\phi^2 - q^2 \right)}{4 \sqrt{\vert q^2 \vert}} \right] 
m_c^3 f_\phi^\perp m_\phi^2 \, \phi_4^\perp(u) \nonumber\\
&+& \left[ - \frac{u \sqrt{\lambda}}{4 \sqrt{\vert q^2 \vert}} \mp \frac{u \left(m_{D_s^\ast}^2 - m_\phi^2 - q^2 \right) + 2 q^2}{4 \sqrt{\vert q^2 \vert}} \right] 
m_c^2 f_\phi^\parallel m_\phi^3 \, \phi_5^\perp(u) \nonumber\\
&+& \left[ - \frac{\sqrt{\lambda} \left[ 2 m_c^2 + u \left(m_{D_s^\ast}^2 - m_\phi^2 - q^2 \right) + 2 q^2\right]}{16 \sqrt{\vert q^2 \vert}} 
\mp \frac{u \lambda}{16 \sqrt{\vert q^2 \vert}} \right] \nonumber\\ 
&~& \cdot f_\phi^\parallel m_\phi^3 \, \tilde{\psi}_5^\perp(u) \nonumber\\
&+& \left[ \frac{\sqrt{\lambda}}{4 \sqrt{\vert q^2 \vert}} \mp \frac{\left( m_{D_s^\ast}^2 - m_\phi^2 -q^2 \right)}{4 \sqrt{\vert q^2 \vert}} \right] 
f_\phi^\parallel m_\phi^3 m_c^2 \left( {\bar \phi}_4^\parallel(u) - {\bar \phi}_5^\perp(u) \right) \,.
\label{eq:Im-OPE-LT-n3} 
\eeq

The imaginary parts of OPE invariant amplitudes with transversal leptonic and longitudinal $\phi$ currents are 
\beq
&~&\frac{1}{\pi} \,{\rm Im} F^{{\rm OPE}}_{1,{\bf \pm 0}}(q^2<0, u) \nonumber\\
&=& \mp f_\phi^\perp m_c m_\phi \,  \phi_2^\perp(u) \mp f_\phi^\parallel m_\phi^2 \left( {\bar \phi}_2^\parallel(u) - {\bar \phi}_3^\perp(u) \right) \nonumber\\
&+& \left[ \sqrt{\lambda} \mp \left( m_{D_s^\ast}^2 - m_\phi^2 - q^2 + 2u m_\phi^2 \right) \right] \frac{f_\phi^\parallel}{2} \, \phi_3^\perp(u) \,,
\label{eq:Im-OPE-TL-n1} \\
&~&\frac{1}{\pi} \,{\rm Im} F^{{\rm OPE}}_{2,{\bf \pm 0}}(q^2<0, u) \nonumber\\
&=& - \left[ \sqrt{\lambda} \left( m_{D_s^\ast}^2 - m_\phi^2 - q^2 + 2u m_\phi^2 \right) \mp \lambda \right] \nonumber\\
&~& \cdot \frac{f_\phi^\parallel}{2} \left( {\bar \phi}_2^\parallel(u) - {\bar \phi}_3^\perp(u) \right) \nonumber\\
&-& \left[ \sqrt{\lambda} \mp \left( m_{D_s^\ast}^2 - m_\phi^2 - q^2 + 2u m_\phi^2 \right) \right] \frac{f_\phi^\parallel m_\phi^2}{8} \, \phi_5^\perp(u) \nonumber\\ 
&-& 2 f_\phi^\parallel m_\phi^2 \, \sqrt{\lambda} \, 
\left(\overset{=}{\psi^\parallel_4}(u) + \overset{=}{\phi^\parallel_2}(u) - 2 \overset{=}{\phi^\perp_3}(u) \right) \nonumber\\
&\mp & f_\phi^\perp m_c m_\phi^3 \,\left(\overset{=}{\psi^\perp_4}(u) + \overset{=}{\phi^\perp_2}(u) - 2 \overset{=}{\phi^\parallel_3}(u) \right) \nonumber\\
&\pm& \left( m_{D_s^\ast}^2 - m_\phi^2 - q^2 + 2u m_\phi^2 \right)  \frac{f_\phi^\perp m_c m_\phi }{2} \left( {\bar \psi}_4^\perp(u) - {\bar \phi}_2^\perp(u) \right) \nonumber\\
&+& \frac{f_\phi^\perp m_\phi m_c }{2} \, \sqrt{\lambda}\, \tilde{\psi}_3^\parallel(u) 
\pm \frac{f_\phi^\parallel m_\phi^4 }{4} \left( {\bar \phi}_4^\parallel(u) - {\bar \phi}_5^\perp(u) \right) \,,
\label{eq:Im-OPE-TL-n2} \\
&~&\frac{1}{\pi} \,{\rm Im} F^{{\rm OPE}}_{3,{\bf \pm 0}}(q^2<0, u) \nonumber\\
&=& \left[ \sqrt{\lambda} \left( m_{D_s^\ast}^2 - m_\phi^2 - q^2 + 2u m_\phi^2 \right) \mp \lambda \pm 2m_c^2m_\phi^2 \right]  \nonumber\\
&~& \cdot \frac{f_\phi^\parallel m_\phi^2}{4} \left( {\bar \phi}_4^\parallel(u) - {\bar \phi}_5^\perp(u) \right) \nonumber\\
&-& 2 f_\phi^\parallel m_\phi^2 \, \sqrt{\lambda} \left( u m_{D_s^\ast}^2 + {\bar u} q^2 - u {\bar u} m_\phi^2 \right) \nonumber\\
&~& \cdot \left( \overset{=}{\psi^\parallel_4}(u) + \overset{=}{\phi^\parallel_2}(u) - 2 \overset{=}{\phi^\perp_3}(u) \right) \nonumber\\
&\pm& \frac{f_\phi^\perp m_c^3 m_\phi^3}{2} \phi_4^\perp(u) 
\pm f_\phi^\perp m_\phi m_c  \, \lambda \left( \overset{=}{\psi^\perp_4}(u) + \overset{=}{\phi^\perp_2}(u) - 2 \overset{=}{\phi^\parallel_3}(u) \right) \nonumber\\
&-& \left[ \sqrt{\lambda} \mp \left( m_{D_s^\ast}^2 - m_\phi^2 - q^2 + 2u m_\phi^2 \right) \right] \frac{ f_\phi^\parallel m_\phi^2 m_c^2 }{4}\, \phi_5^\perp(u) \,.
\label{eq:Im-OPE-TL-n3} 
\eeq

For the transversal helicity form factor, the imaginary parts are
\beq
&~&\frac{1}{\pi} \,{\rm Im} F^{{\rm OPE}}_{1,{\bf \pm\mp}}(q^2<0, u) \nonumber\\
&=& \left[ \frac{{\bar u} \sqrt{\lambda}}{2 m_{D_s^\ast}} \pm \frac{\left( (1+u) m_{D_s^\ast}^2 - {\bar u} m_\phi^2 + {\bar u} q^2 \right)}{2 m_{D_s^\ast}} \right]  
f_\phi^\parallel m_\phi \, \phi_3^\perp(u) \nonumber\\
&+& \left[ \frac{\sqrt{\lambda}}{2 m_{D_s^\ast}} \pm \frac{\left( m_{D_s^\ast}^2+m_\phi^2-q^2 \right)}{2 m_{D_s^\ast}} \right]  \nonumber\\
&~& \cdot \Big[ f_\phi^\perp m_c \, \phi_2^\perp(u) + f_\phi^\perp m_\phi \left( {\bar \phi}_2^\parallel(u) - {\bar \phi}_3^\perp(u) \right) \Big] \nonumber\\
&+& \frac{\sqrt{\lambda}}{4 m_{D_s^\ast}} \, f_\phi^\parallel m_\phi \, \tilde{\psi}_3^\perp(u) \,,
\label{eq:Im-OPE-T-n1} 
\eeq
\beq
&~&\frac{1}{\pi} \,{\rm Im} F^{{\rm OPE}}_{2,{\bf \pm\mp}}(q^2<0, u) \nonumber\\
&=& \left[ \frac{ \left( (1+u) m_{D_s^\ast}^2 - {\bar u} m_\phi^2 + {\bar u} q^2 \right) \sqrt{\lambda} }{2 m_{D_s^\ast}} \pm \frac{{\bar u} \lambda}{2 m_{D_s^\ast}} \right] \nonumber\\
&~& \cdot \Big[ \frac{f_\phi^\parallel m_\phi \, \tilde{\psi}_3^\perp(u)}{4} \mp \frac{f_\phi^\parallel m_\phi^3 \, \phi_5^\perp(u) }{4\sqrt{\lambda}} \nonumber\\
&~& \; \mp \frac{f_\phi^\perp m_\phi^2 m_c \, \left( {\bar \psi}_4^\perp(u) - {\bar \phi}_2^\perp(u) \right)}{\sqrt{\lambda}}\Big] \nonumber\\
&+& \left[ \frac{\sqrt{\lambda}}{2 m_{D_s^\ast}} \pm \frac{\left( m_{D_s^\ast}^2+m_\phi^2-q^2 \right)}{2 m_{D_s^\ast}} \right] \nonumber\\
&~& \cdot \Big[ f_\phi^\perp m_\phi^2 m_c  \left( \overset{=}{\psi^\perp_4}(u) + \overset{=}{\phi^\perp_2}(u) - 2 \overset{=}{\phi^\parallel_3}(u) \right) \nonumber\\
&~& \; - \frac{f_\phi^\parallel m_\phi^3 \left( {\bar \phi}_4^\parallel(u) - {\bar \phi}_5^\perp(u) \right) }{4} \Big] \nonumber\\ 
&-& \frac{{\bar u} \sqrt{\lambda}}{m_{D_s^\ast}} \, f_\phi^\parallel m_\phi^3 
\left( \overset{=}{\psi^\parallel_4}(u) + \overset{=}{\phi^\parallel_2}(u) - 2 \overset{=}{\phi^\perp_3}(u) \right) \nonumber\\
&-& \frac{\sqrt{\lambda}}{16 m_{D_s^\ast}} \, f_\phi^\parallel m_\phi^3 \, \tilde{\psi}_5^\perp(u)  \,,
\label{eq:Im-OPE-T-n2} \\
&~&\frac{1}{\pi} \,{\rm Im} F^{{\rm OPE}}_{3,{\bf \pm\mp}}(q^2<0, u) \nonumber\\
&=& \left[ - \frac{{\bar u} \sqrt{\lambda}}{4 m_{D_s^\ast}}  \mp \frac{\left( (1+u) m_{D_s^\ast}^2 - {\bar u} m_\phi^2 + {\bar u} q^2 \right)}{4 m_{D_s^\ast}} \right] 
f_\phi^\parallel m_\phi^3 m_c^2 \, \phi_5^\perp(u) \nonumber\\
&-& \left[ \frac{\left( 2 m_c^2 + (1+u) m_{D_s^\ast}^2 - {\bar u} m_\phi^2 + {\bar u} q^2 \right) \sqrt{\lambda}}{16 m_{D_s^\ast}} 
\pm \frac{{\bar u} \lambda}{16 m_{D_s^\ast}} \right] \nonumber\\ 
&~& \cdot f_\phi^\parallel m_\phi^3 \, \tilde{\psi}_5^\perp(u) \nonumber\\
&-& \left[ \frac{\sqrt{\lambda}}{4 m_{D_s^\ast}} \pm \frac{\left(m_{D_s^\ast}^2+m_\phi^2-q^2 \right)}{4 m_{D_s^\ast}} \right] \nonumber\\
&~& \cdot \Big[ f_\phi^\perp m_\phi^2 m_c^3 \, \phi_4^\perp(u) + f_\phi^\parallel m_\phi^3 m_c^2 \left( {\bar \phi}_4^\parallel(u) - {\bar \phi}_5^\perp(u) \right) \Big] \,.
\label{eq:Im-OPE-T-n3} 
\eeq

\end{appendix}

\end{document}